\documentclass[twocolumn, preprint]{aastex63}
\usepackage{amsmath,amstext}
\usepackage[T1]{fontenc}
\usepackage{apjfonts} 
\usepackage{natbib}
\usepackage{microtype}
\usepackage{longtable}
\usepackage{verbatim}
\usepackage{upgreek}
\usepackage{multirow}
\usepackage{hyperref}
\received{May 16, 2022}
\revised{XXX}
\accepted{XXX}
\submitjournal{ApJ}

\shorttitle{Evolution of Spatially Resolved Star Formation between $0.5\lesssim z \lesssim1.7$}
\shortauthors{Matharu et al.}
\graphicspath{{./}{figures/}}

\begin{document}

\title{CLEAR: The Evolution of Spatially Resolved Star Formation in Galaxies between $0.5\lesssim z \lesssim1.7$ using H$\upalpha$ Emission Line Maps}

\correspondingauthor{Jasleen Matharu}
\email{jmatharu@tamu.edu}

\author[0000-0002-7547-3385]{Jasleen Matharu}
\affiliation{Department of Physics and Astronomy, Texas A\&M University, College Station, TX, 77843-4242, USA\\}
\affiliation{George P.\ and Cynthia Woods Mitchell Institute for
Fundamental Physics and Astronomy, Texas A\&M University, College
Station, TX, 77845-4242, USA\\}

\author[0000-0001-7503-8482]{Casey Papovich}
\affiliation{Department of Physics and Astronomy, Texas A\&M University, College Station, TX, 77843-4242, USA\\}
\affiliation{George P.\ and Cynthia Woods Mitchell Institute for
Fundamental Physics and Astronomy, Texas A\&M University, College
Station, TX, 77845-4242, USA\\}

\author[0000-0002-6386-7299]{Raymond C. Simons}
\affil{Space Telescope Science Institute, 3700 San Martin Drive,
  Baltimore, MD, 21218 USA}

\author[0000-0003-1665-2073]{Ivelina Momcheva}
\affil{Space Telescope Science Institute, 3700 San Martin Drive,
  Baltimore, MD, 21218 USA}

\author[0000-0003-2680-005X]{Gabriel Brammer}
\affiliation{Cosmic Dawn Center, Niels Bohr Institute, University of Copenhagen, Jagtvej 128, 2200 Copenhagen N, Denmark\\}

\author[0000-0001-7673-2257]{Zhiyuan Ji}
\affiliation{University of Massachusetts Amherst, 710 North Pleasant Street, Amherst, MA 01003-9305, USA}

\author[0000-0001-8534-7502]{Bren E. Backhaus } 
\affil{Department of Physics, University of Connecticut, Storrs, CT 06269, USA}

\author[0000-0001-7151-009X]{Nikko J. Cleri}
\affiliation{Department of Physics and Astronomy, Texas A\&M University, College Station, TX, 77843-4242, USA\\}
\affiliation{George P.\ and Cynthia Woods Mitchell Institute for
Fundamental Physics and Astronomy, Texas A\&M University, College
Station, TX, 77845-4242, USA\\}

\author[0000-0001-8489-2349]{Vicente Estrada-Carpenter}
\affiliation{Department of Physics and Astronomy, Texas A\&M University, College Station, TX, 77843-4242, USA\\}
\affiliation{George P.\ and Cynthia Woods Mitchell Institute for
Fundamental Physics and Astronomy, Texas A\&M University, College
Station, TX, 77845-4242, USA\\}
\affiliation{Department of Astronomy \& Physics, Saint Mary's University, 923 Robie Street, Halifax, NS, B3H 3C3, Canada}

\author[0000-0001-8519-1130]{Steven L. Finkelstein}
\affil{Department of Astronomy, The University of Texas at Austin, Austin, TX, 78759 USA}

\author[0000-0002-0496-1656]{Kristian Finlator}
\affiliation{Cosmic Dawn Center, Niels Bohr Institute, University of Copenhagen, Jagtvej 128, 2200 Copenhagen N, Denmark\\}
\affil{Department of Astronomy, New Mexico State University, Las Cruces, NM 88003, USA\\}

\author[0000-0002-7831-8751]{Mauro Giavalisco}
\affil{Astronomy Department, University of Massachusetts, Amherst, MA, 01003 USA} 

\author[0000-0003-1187-4240]{Intae Jung}
\affil{Department of Physics, The Catholic University of America, Washington, DC, 20064 USA}
\affil{Astrophysics Science Division, Goddard Space Flight Center, Greenbelt, MD, 20771 USA}

\author[0000-0002-9330-9108]{Adam Muzzin}
\affiliation{Department of Physics and Astronomy, York University, 4700 Keele Street, Toronto, ON, M3J 1P3, Canada\\}

\author[0000-0003-1065-9274]{Annalisa Pillepich}
\affiliation{Max-Planck-Institut f\"ur Astronomie, K\"oniggstuhl 17, D-69117 Heidelberg, Germany\\}

\author[0000-0002-1410-0470]{Jonathan R. Trump}
\affil{Department of Physics, 196A Auditorium Road Unit 3046, University of Connecticut, Storrs, CT 06269 USA}

\author[0000-0001-6065-7483]{Benjamin Weiner}
\affil{MMT/Steward Observatory, 933 N. Cherry St., University of Arizona, Tucson,
AZ 85721, USA}



\def\editR{\textbf}
\def\ha{H$\upalpha$}

\begin{abstract}
Using spatially resolved H$\upalpha$ emission line maps of star-forming galaxies, we study the evolution of gradients in galaxy assembly over a wide range in redshift ($0.5\lesssim z \lesssim1.7$). Our $z\sim0.5$ measurements come from deep {\it Hubble Space Telescope} WFC3 G102 grism spectroscopy obtained as part of the CANDELS Lyman-$\upalpha$ Emission at Reionization (CLEAR) Experiment. For star-forming galaxies with Log$(M_{*}/\mathrm{M}_{\odot})\geqslant8.96$, the mean H$\upalpha$ effective radius is $1.2\pm0.1$ times larger than that of the stellar continuum, implying inside-out growth via star formation. This measurement agrees within $1\upsigma$ with those measured at $z\sim1$ and $z\sim1.7$ from the 3D-HST and KMOS$^{\mathrm{3D}}$ surveys respectively, implying no redshift evolution. However, we observe redshift evolution in the stellar mass surface density within 1 kiloparsec ($\Sigma_\mathrm{1kpc}$). Star-forming galaxies at $z\sim0.5$ with a stellar mass of Log$(M_{*}/\mathrm{M}_{\odot})=9.5$ have a ratio of $\Sigma_\mathrm{1kpc}$ in H$\upalpha$ relative to their stellar continuum that is lower by $(19\pm2)\%$ compared to $z\sim1$ galaxies. $\Sigma_{1\mathrm{kpc, H}\upalpha}$/$\Sigma_{1\mathrm{kpc,Cont}}$  decreases towards higher stellar masses. The majority of the redshift evolution in $\Sigma_{1\mathrm{kpc, H}\upalpha}$/$\Sigma_{1\mathrm{kpc,Cont}}$ versus stellar mass stems from the fact that Log($\Sigma_{1\mathrm{kpc, H}\upalpha}$) declines twice as much as Log($\Sigma_{1\mathrm{kpc, Cont}}$) from $z\sim 1$ to 0.5 (at a fixed stellar mass of Log$(M_{*}/\mathrm{M}_{\odot})=9.5$). By comparing our results to the TNG50 cosmological magneto-hydrodynamical simulation, we rule out dust as the driver of this evolution. Our results are consistent with inside-out quenching following in the wake of inside-out growth, the former of which drives the significant drop in $\Sigma_{1\mathrm{kpc, H}\upalpha}$ from $z\sim1$ to $z\sim0.5$.


\end{abstract}



\section{Introduction}
\label{sec:intro}

In a Lambda Cold Dark Matter ($\Lambda$CDM) Universe such as ours, the assembly of galaxies is thought to be controlled by the nature of the dark matter haloes within which galaxies reside. The mass of the dark matter halo dictates the rate of gas accretion from the cosmic web and its angular momentum distribution the radial distribution of stars \citep{White&Rees, Fall1980, Dalcanton1997, VanDenBosch2001, Dekel2013}. In a simple model of galaxy formation, the sizes of galaxy disks are assumed proportional to the sizes of their dark matter haloes \citep{Mo1998}. As haloes grow in size with time, star formation in galaxies should progress at larger galactocentric radii. More recently however, sophisticated hydrodynamic cosmological simulations have revealed the assembly of galaxies is a delicate balance between multiple physical processes, complicating this simple picture \citep{Keres2005,Brooks2009,Sales2012,Ubler2014,Genel2015,Minchev2015, Nelson2015d,Fielding2017,Angles-Alcazar2017,Sparre2017,Beckmann2017,Mitchell2018,Correa2018,Dawoodbhoy2018,Peeples2019,Nelson2019,KarChowdhury2020,Nelson2020,Mitchell2020,Wright2020,Bennett2020,Watts2020,Okalidis2021,Wright2021,Mitchell2021}.

An important confirmation of this picture is evidence that galaxies grow from the ``inside-out''. Observational studies on the integrated star formation in galaxies over a wide range in redshift ($0<z<8$) have revealed the rise ($z\sim8$ to $z\sim2$) and fall ($z\sim2$ to $z=0$) of the amount of star formation per unit solar mass per unit time per unit volume (\citealt{Madau&Dickinson2014} and references therein). Whilst able to provide indirect evidence for star formation and stellar mass build-up in galaxies, these studies are insufficient in helping us definitively answer {\it how} star formation proceeds in galaxies. Answering this question requires spatially resolved observations of galaxies that are able to show us where the star formation process starts and finishes in a galaxy. 

Some of the first observational campaigns in pursuit of this were naturally confined to the local Universe where it is easier to spatially resolve galaxies. Narrow band imaging over wavelengths sensitive to ongoing star formation over different timescales were used \citep{Hodge1983,Athanassoula1993,Ryder1994,Kenney1999,Koopmann2004a,Koopmann2004,Koopmann2006,Cortes2006,Crowl2006,MunozMateos2007,Abramson2011a,Vollmer2012,Gavazzi2013,Kenney2015,Abramson2016b,Lee2017,Gavazzi2018,Cramer2019,Boselli2020}. The presence of \ha~emission in galaxy spectra indicates the presence of massive, young O and B stars that emit ultraviolet radiation \citep{Kennicutt1998}. With lifetimes of approximately 10 Myr, \ha~emission due to these stars allows us to trace star formation on very short timescales. By comparing the sizes and morphologies of galaxy images in \ha~versus the rest-frame optical which is a tracer of predominantly older stars (the ``stellar continuum"), we can compare {\it where} star formation in the galaxy has been taking place in the past 10 Myr versus where it had occurred in the distant past.

The first direct evidence for inside-out growth via star formation in galaxies at high redshift was provided at $z\sim1$ as part of the 3D-HST Survey \citep{VanDokkum2011,Brammer2012,Momcheva2016}. Using spatially resolved space-based slitless spectroscopy with the {\it Wide Field Camera 3} (WFC3) on-board the {\it Hubble Space Telescope} (HST),  \cite{Nelson2012,Nelson2015} were able to show that ongoing star formation traced by \ha~emission occurs in disks that are more extended than those occupied by existing stars in star-forming galaxies. Subsequently, a similar study was conducted at $z\sim1.7$ using ground-based integral field spectroscopy with the {K-band Multi-Object Spectrograph} (KMOS) on the {\it Very Large Telescope} (VLT) as part of the KMOS$^{\mathrm{{3D}}}$ survey \citep{Wisnioski2015,Wisnioski2019} confirming consistent, albeit marginally more extended star-forming disks as those observed at $z\sim1$ \citep{Wilman2020}. In the local Universe ($z=0$), there is some tentative evidence for inside-out growth (e.g. \citealt{MunozMateos2007}) but the star-forming disk has been found to have the same spatial extent as the stellar disk \citep{James2009,Fossati2013}. These handful of results tentatively suggest inside-out growth via star formation slows down with the decline in cosmic star formation.

Understanding the evolution in the cosmic star formation history requires us to understand how star formation and the quenching of star formation operate both in the low and high redshift Universe. Tracking spatially resolved star formation during the epoch of cosmic star formation decline ($0<z<2$) could provide direct observational evidence for the dominant physical mechanism responsible for the quenching of star formation. Recently, \cite{Matharu2021a} demonstrated how the same technique of using HST WFC3 space-based slitless spectroscopy in \cite{Nelson2015} can be used to provide direct evidence for rapid outside-in quenching due to ram-pressure stripping in $z\sim1$ galaxy clusters. Other similar measurements not using space-based slitless spectroscopy have been similarly successful in providing direct evidence of ram-pressure stripping in galaxy clusters at lower redshifts \citep{Koopmann2006,Bamford2007,Jaffe2011,Bosch2013,Vulcani2016,Finn2018,Vaughan2020}, with evidence of its sub-dominance in a $z=2.5$ protocluster \citep{Suzuki2019}. Similarly, making star-forming versus stellar disk measurements for the general star-forming galaxy population at multiple epochs between $0<z<2$ can help us better constrain the physical mechanism driving quenching in the Universe. 

Over the past decade, there has been growing observational evidence that the way in which quenching operates alters between $0.5\lesssim z \lesssim1.5$. Out to $z\sim1$, quenching due to internal processes in the galaxy (``self-quenching" or ``mass quenching") and external processes due to the environment within which galaxies reside (``environment quenching") are completely separable \citep{Peng2010d,Muzzin2012}. However, the efficiency of environment quenching starts to become an increasing function of stellar mass at $z \gtrsim 1$ \citep{Kawinwanichakij2017,Papovich2018,VanderBurg2020}. Investigating the nature of spatially resolved star formation between $0<z<1$ and placing it within the context of the $z=0$, $z\sim1$ and $z\sim1.7$ measurements could therefore be crucial in determining the process responsible for this turning point in the physics of quenching.

In this paper, we use deep HST WFC3 spatially resolved space-based slitless spectroscopy from the CANDELS Lyman-$\upalpha$ Emission at Reionization (CLEAR) Survey to measure spatially resolved star formation at an intermediate redshift of $z\sim0.5$, a crucial epoch for understanding the turning point in the physics of quenching. We deliberately follow the methodology of \cite{Nelson2015} to reduce systematics in our comparison to their $z\sim1$ results and also compare to the $z\sim1.7$ KMOS$^{\mathrm{{3D}}}$ results from \cite{Wilman2020}.

This paper is organized as follows. Section~\ref{sec:data} describes the three data sets used in this study. In Section~\ref{methodology} we describe the grism data reduction, sample selection and stacking procedures. The size determination process for the H$\upalpha$ and continuum maps obtained from CLEAR is described in Section~\ref{size_determination}. We describe the morphology parameters we explore in Section~\ref{morph_parameters}. In Section~\ref{sec:results}, we present our results. We then discuss their physical implications in Section~\ref{discussion} by comparing to the outcome of cosmological galaxy simulations. Finally, we summarize our findings in Section~\ref{summary}.

All magnitudes quoted are in the AB system, logarithms are in base 10, effective radii are not circularized and we assume a $\Lambda$CDM cosmology with $\Omega_{m}=0.307$, $\Omega_{\Lambda}=0.693$ and $H_{0}=67.7$~kms$^{-1}$~Mpc$^{-1}$ \citep{Planck2015}.

\section{Data}
\label{sec:data}

In this section, we describe the three data sets we use to probe \ha~sizes and morphologies of star-forming galaxies over a wide range in stellar mass and redshift. This includes measurements from CLEAR (Section~\ref{CLEAR}) at $0.22 \lesssim z \lesssim 0.75$, 3D-HST \citep{Nelson2015} at $0.7<z<1.5$ (Section~\ref{3dhst}) and KMOS$^{\mathrm{{3D}}}$ \citep{Wilman2020}  at $0.7 \lesssim z \lesssim 2.7$ (Section~\ref{kmos3d}).

\subsection{The CLEAR Survey}
\label{CLEAR}

The CANDELS Lyman-$\upalpha$ Emission at Reionization (CLEAR) Experiment (GO-14227, PI: Papovich) is a deep {\it Hubble Space Telescope} (HST), {\it Wide Field Camera 3} (WFC3) F105W/G102 direct imaging and grism spectroscopy survey, covering 12 pointings across the GOODS-S and GOODS-N deep CANDELS fields \citep{Grogin2011,Koekemoer2011}. The 6 pointings in GOODS-S have a maximum of 12 orbit depth, whilst the 6 pointings in GOODS-N have a maximum of 12 orbit depth, of which two orbits are obtained from GO-13420 (PI: Barro). Each pointing is observed with at least 3 orients separated by more than 10 degrees, allowing for a reduction in source confusion and improved mitigation of grism spectra contamination. There is also archival F105W/G102 observations covering some CLEAR pointings in GOODS-S from GO/DD-11359 (PI: O'Connell) and GO-13779 (PI: Malhotra), the latter of which is an additional 40 orbits on a CLEAR pointing covering the Hubble Ultra Deep Field (The Faint Infrared Grism Survey, \citealt{Pirzkal2017}). A full description of the CLEAR Survey will be discussed in an upcoming survey paper (Simons et al., in prep).

The CLEAR survey was primarily designed to measure the evolution in the strength of Lyman-$\upalpha$ emission from galaxies at $6.5\leqslant z \leqslant 8.2$ \citep{Jung2021}. By virtue of being a spectroscopic survey in the well-studied GOODS-S and GOODS-N fields, the CLEAR data has high utility, especially when combined with the wealth of ancillary photometry spanning from the ultraviolet to the near-infrared \citep{Skelton2014a} and WFC3/G141 grism spectroscopy from the 3D-HST survey (see Section~\ref{3dhst}). So far, it has allowed for studies on the ages, metallicities and star formation histories of high redshift massive quiescent galaxies \citep{Estrada-Carpenter2018,Estrada-Carpenter2020}, the gas-phase metallicity gradients of star-forming galaxies between $0.6 < z < 2.6$ \citep{Simons2020a}, emission line ratios at $z\sim1.5$ \citep{Backhaus2022}, an exploration of Paschen-$\upbeta$ as a star formation rate indicator in the local ($z\lesssim0.3$) Universe \citep{Cleri2022} and the ionization and chemical enrichment properties of star-forming galaxies at $1.1 \lesssim z \lesssim 2.3$ \citep{Papovich2022}.

The WFC3 F105W direct imaging in CLEAR provides high spatial resolution (FWHM$\sim0.128^{\prime\prime}$) rest-frame optical imaging of the stellar continuum for galaxies between $0.22 \lesssim z \lesssim 0.75$. The WFC3 G102 grism spectroscopy provides low spectral resolution ($R\sim210$ at $10000$~{\AA}) but high spatial resolution (FWHM$\sim0.128^{\prime\prime}$) two-dimensional spectra in the wavelength range \mbox{$8000<~\lambda~/$~{\AA}~$<11500$}. G102 grism spectroscopy effectively provides an image of the galaxy at specific wavelengths in $24.5${\AA} increments spanning the wavelength range of the grism. An unresolved\footnote{Typical kinematic widths of star-forming galaxies are unresolved by the G102 and G141 grisms ($\lesssim 800\mathrm{km~s}^{-1}$ and $\lesssim 1000\mathrm{km~s}^{-1}$ per pixel, respectively \citep{Outini2019}).} emission line in this two-dimensional spectrum manifests itself as an image of the galaxy in that line on top of the underlying stellar continuum. The combination of high spatial resolution of the WFC3 and low spectral resolution of the G102 grism allows for spatially-resolved emission line maps of galaxies. With the G102 grism, we can obtain spatially-resolved H$\upalpha$ emission line maps of galaxies between $0.22 \lesssim z \lesssim 0.75$. 

For the study presented in this paper, we will use the H$\upalpha$ emission line maps obtained from the G102 grism observations in conjunction with the F105W direct imaging from CLEAR to study how star formation proceeds spatially in $z\sim0.5$ star-forming galaxies.

\subsection{The 3D-HST Survey}
\label{3dhst}

The 3D-HST survey was a 248-orbit near-infrared spectroscopic program with HST. The survey covered three quarters of the CANDELS treasury survey area with WFC3 F140W/G141 direct imaging and grism spectroscopy to 2-orbit depth \citep{VanDokkum2011,Brammer2012,Momcheva2016}. The WFC3 F140W direct imaging in 3D-HST provides high spatial resolution (FWHM$\sim0.141^{\prime\prime}$) rest-frame optical imaging of the stellar continuum for galaxies between \mbox{$0.7<z<1.5$}. The WFC3 G141 grism spectroscopy provides low spectral resolution ($R\sim130$ at $14000$~{\AA}) but high spatial resolution (FWHM$\sim0.141^{\prime\prime}$) two-dimensional spectra in the wavelength range $10750\leqslant~\lambda~/$~{\AA}~$\leqslant17000$. G141 grism spectroscopy effectively provides an image of the galaxy at specific wavelengths in $46.5${\AA} increments spanning the wavelength range of the grism. An unresolved emission line in this two-dimensional spectrum manifests itself as an image of the galaxy in that line on top of the underlying continuum.

The combination of high spatial resolution of the WFC3 and low spectral resolution of the G141 grism allows for spatially-resolved emission line maps of galaxies. With the G141 grism, we can obtain spatially-resolved H$\upalpha$ emission line maps of galaxies between \mbox{$0.7<z<1.5$}. \cite{Nelson2015} used the H$\upalpha$ emission line maps in conjunction with the F140W direct imaging from 3D-HST to study how star formation proceeds spatially in $z\sim1$ star-forming galaxies. We closely follow the methodology in \cite{Nelson2015} -- and in places expand upon it -- for our analogous study at $0.22 \lesssim z \lesssim 0.75$ with the CLEAR dataset (Section~\ref{CLEAR}). We compare our results to those obtained by 3D-HST in Section~\ref{evolution}.

\subsection{The KMOS$^{\mathrm{{\it 3D}}}$ Survey}
\label{kmos3d}

The KMOS$^{\mathrm{{3D}}}$ survey \citep{Wisnioski2015,Wisnioski2019} was a near-infrared integral field spectroscopy survey with the K-band Multi-Object Spectrograph (KMOS) on the ESO Very Large Telescope (VLT). The survey targeted galaxies in the 3D-HST (Section~\ref{3dhst}) and CANDELS surveys that were accessible from Paranal Observatory. Target galaxies had a \mbox{$K_{s}$ magnitude < 23}, a spectroscopic or grism redshift indicating the spectrum in the vicinity of the H$\upalpha$ line would be somewhat free of atmospheric OH lines and visible in the KMOS $YJ$, $H$ or $K$-bands.

The survey led to a sample of 645 galaxies in the redshift range $0.7 \lesssim z \lesssim 2.7$ with moderate spatial resolution (median FWHM$\sim0.456^{\prime\prime}$) H$\upalpha$ emission line maps. \cite{Wilman2020} used these H$\upalpha$ emission line maps in conjunction with HST WFC3 F125W/F160W direct imaging from 3D-HST+CANDELS \citep{Skelton2014a} to study how star formation proceeds spatially in 281 $z\sim1.7$ star-forming galaxies. We compare our results from the CLEAR dataset (Section~\ref{CLEAR}) at $z\sim0.5$ to those obtained by KMOS$^{\mathrm{{3D}}}$ in Section~\ref{evolution}.

\section{Data Reduction, Sample Selection \& Stacking}
\label{methodology}

\subsection{Grism Data Reduction}
\label{data_reduction}
In this section we summarize the grism data reduction process for the CLEAR dataset (Section~\ref{CLEAR}). For a detailed explanation of the full data reduction process, we refer the reader to Section 2 of \cite{Simons2020a}.

The newest version of the {\it Grism redshift \& line analysis software for space-based slitless spectroscopy} (\texttt{Grizli}\footnote{\url{https://grizli.readthedocs.io/en/master/}}, \citealt{Grizli2021}) is used to reduce the G$102$ data and create the H$\upalpha$ emission line maps. The reduction for the CLEAR dataset includes the reduction of overlapping F105W, G102 and G141 data from other archival programs when available. \texttt{Grizli} starts by querying the Mikulski Archive for Space Telescopes (MAST) for all available HST WFC3 observations on and in the vicinity of the CLEAR survey footprint. \texttt{Grizli} is designed to fully process HST imaging and grism spectroscopy datasets, from the retrieval and pre-processing of raw observations to the extraction of one-dimensional (1D) and two-dimensional (2D) spectra leading to the generation of emission line maps.

\subsubsection{Pre-processing}

\texttt{Grizli} processes the raw WFC3 data using the \texttt{calwf3} pipeline, including corrections for variable sky backgrounds \citep{Brammer2016}, identification of hot pixels and cosmic rays \citep{Gonzaga2012}, flat-fielding, sky subtraction \citep{Brammer2015}, astrometric corrections and alignment.

\subsubsection{Source Extraction, Contamination Modeling and Subtraction}
\label{se_cont_sub_grizli}
A contamination model for each HST pointing is created by forward-modelling the HST F105W full-field mosaic and it is this model that is used to subtract contaminating grism spectra for each grism spectrum. The contamination model is created in two steps. The first step is a flat model for every source with F105W magnitude < 25. Then third-order polynomial models are created for all sources with F105W magnitude < 24, subtracting the initial flat models.

All sources in the field of view with F105W magnitude < 25 have their grism spectra extracted with \texttt{Grizli}. The G102 exposure times for all extracted sources range from  2 -- 30 hours, with median values of 2.5 and 7.7 hours for GOODS-N and GOODS-S respectively. 533 extracted sources only have G102 coverage and 4707 of extracted sources have both G102 and G141 coverage.

\subsubsection{Grism Redshift Determination}
\label{grism_z}
Grism redshifts are determined by fitting the grism spectra and available multiwavelength photometry simultaneously. A basis set of template Flexible Stellar Population Synthesis models (\texttt{FSPS}; \citealt{Conroy2009,Conroy2010}) are projected to the pixel grid of the 2D grism exposures using the spatial morphology from the HST F105W image. The 2D template spectra are then fit to the observed spectra with non-negative least squares. When performing a redshift fit, the user provides a trial redshift range. For the CLEAR extraction, $0<z<12$ was used. The final grism redshift is taken to be where the $\chi^{2}$ is minimized across the grid of trial redshifts.

\subsubsection{Making H$\alpha$ Maps}
\label{making_maps}

As part of the grism redshift determination process (Section~\ref{grism_z}), a combination of line complex and continuum templates are used. Included in the line complex templates are [\ion{O}{2}] + [\ion{Ne}{3}], [\ion{O}{3}] + H$\upbeta$ and \ha~+ [\ion{S}{2}]. The line complexes help to break redshift degeneracies. The full grism redshift determination process (Section~\ref{grism_z}) leads to a full line + continuum model for each 2D grism spectrum. The user is then able to create drizzled\footnote{Raw images from HST are distorted, and are therefore ``drizzled" to some tangent point to create an undistorted image.} continuum-subtracted narrow-band maps at any wavelength. The \ha~emission line map is created by deliberately choosing the wavelength at which it is detected in the G102 grism from the grism redshift determination process. This extraction uses the full World Coordinate System (WCS) information for each individual grism exposure of the source.

Emission line maps are generated by subtracting the best-fit stellar continuum model under the assumption that the F105W direct image represents the morphology of the source in the stellar continuum. For this work, we made \ha~emission line maps covering $189\times189$ pixels with a pixel scale of 0.1$^{\prime\prime}$. These dimensions were chosen such that 2.5 times the elliptical Kron radius (as was done in \citealt{VanderWel2012}) in F105W of the largest galaxy in the sample could be encompassed.


Due to the low spectral resolution of the G102 grism (see Section~\ref{CLEAR}), \ha~and [\ion{N}{2}] are blended. The presence of [\ion{N}{2}] in our \ha~emission line maps serves only to increase the overall flux by approximately one tenth of the \ha~flux (e.g. \citealt{Faisst2018} and see Section~\ref{stacking_procedure}). Our study focuses only on shape and morphology measurements of the \ha~emission. Therefore the results in our study are unlikely to be effected by this blending.

\subsection{Sample Selection} 
\label{sample}

\begin{table}
	\centering
	\caption{Summary of Sample Selection in order}
	\label{tab:sample_selection_table}
	\begin{tabular}{l} 
		\hline
		Sample Selection Criteria \\
		\hline
		F105W magnitude < 25 \\
		H$\upalpha$ flux > 0 \mbox{ergs~s$^{-1}$~cm$^{-2}$} \\
		$0.22 \lesssim z \lesssim 0.75$\\
        Star-forming galaxies: $(U-V)_{\mathrm{rest}} < 0.88~(V-J)_{\mathrm{rest}}+0.64$\\
        AGN detected from X-ray removed \\
		Log$(M_{*}/\mathrm{M}_{\odot}) \geqslant 8.96$ \\
		H$\upalpha$ stack SNR > 48 \\
		All fields and GS4 sample separation (see Section~\ref{all_GS4_separation}) \\
		\hline
	\end{tabular}
\end{table}

\begin{figure*}
	\centering\includegraphics[width=0.88\textwidth]{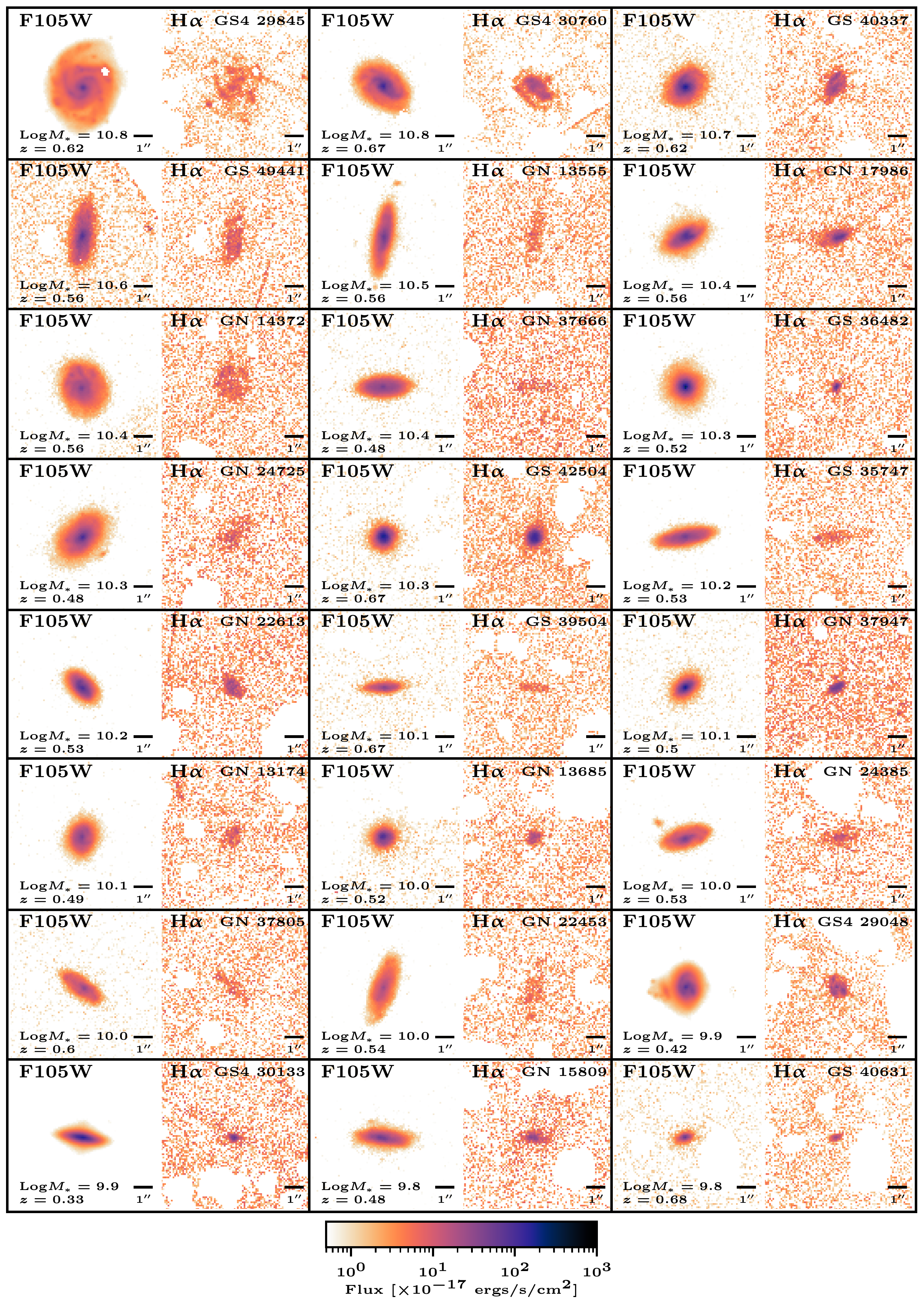}
    \caption{Zoomed-in regions ($89\times89$~pixels, pixel scale = 0.1$^{\prime\prime}$) of {\it individual} stellar continuum and H$\upalpha$ thumbnails for select galaxies in the CLEAR sample ordered by descending stellar mass. Original thumbnails are $189\times189$~pixels (see Section~\ref{making_maps} for reasoning). H$\upalpha$ emission line maps are multiplied by 50 for visibility. The 3D-HST source ID for each galaxy is marked in the top right corner of each H$\upalpha$ thumbnail. GS = GOODS-S, GN = GOODS-N and see Section~\ref{all_GS4_separation} for GS4 explanation. The stellar mass and grism redshift is stated in the bottom left corner of each F105W thumbnail.}
    \label{fig:eyecandy1}
\end{figure*}

\begin{figure*}
	\centering\includegraphics[width=0.88\textwidth]{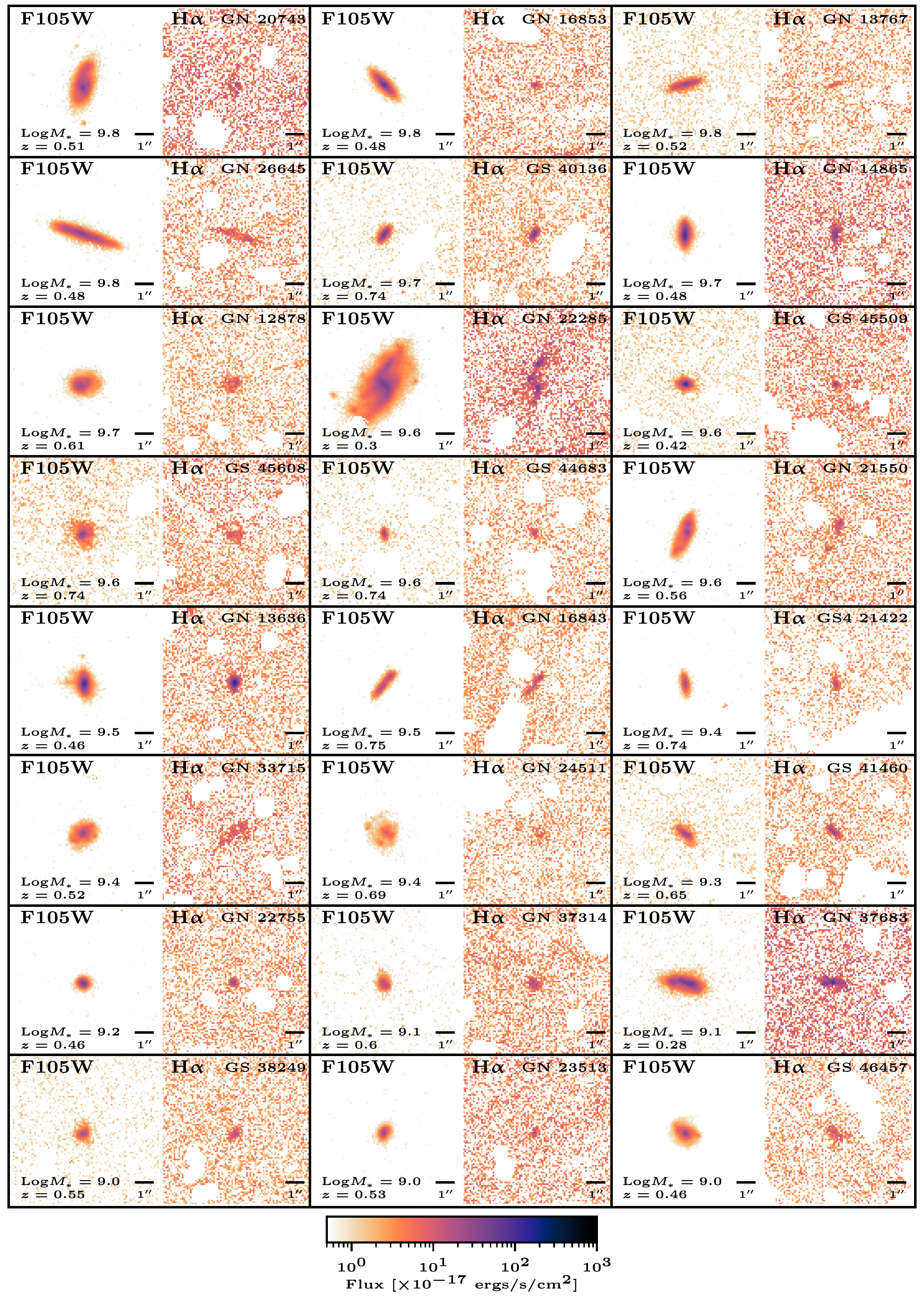}
    \caption{Figure~\ref{fig:eyecandy1} continued}
    \label{fig:eyecandy2}
\end{figure*}

In this section we describe the sample selection for our study. A summary of the sample selection in the order it is applied can be viewed in Table~\ref{tab:sample_selection_table}. A selection of some galaxies in the final sample with their corresponding \ha~emission line maps are shown in Figures~\ref{fig:eyecandy1} and \ref{fig:eyecandy2} to illustrate the nature of our data.

Our sample is first selected to have F105W magnitude < 25 as the limiting magnitude of the \texttt{Grizli} data reduction process (Section~\ref{se_cont_sub_grizli}). Within the sample of extracted grism spectra with \texttt{Grizli}, we find all the galaxies in the CLEAR fields with a \ha~flux > 0 \mbox{ergs~s$^{-1}$~cm$^{-2}$} as detected by \texttt{Grizli} in their G102 spectrum. Then, given the rest-frame wavelength \texttt{Grizli} uses for the \ha~line (6564.61{\AA}), we ensure that the \ha~line has been detected within the high-throughput wavelength range of the G102 grism (\mbox{$8000<~\lambda~/$~{\AA}~$<11500$}, see Section~\ref{CLEAR}), given the grism redshift determined for each galaxy (Section~\ref{grism_z}). This gives a sample of 613 galaxies in GOODS-N and 305 in GOODS-S.

\subsubsection{Selecting Star-forming Galaxies}
\label{select_sf}

\begin{figure*}

	\centering\includegraphics[width=\textwidth]{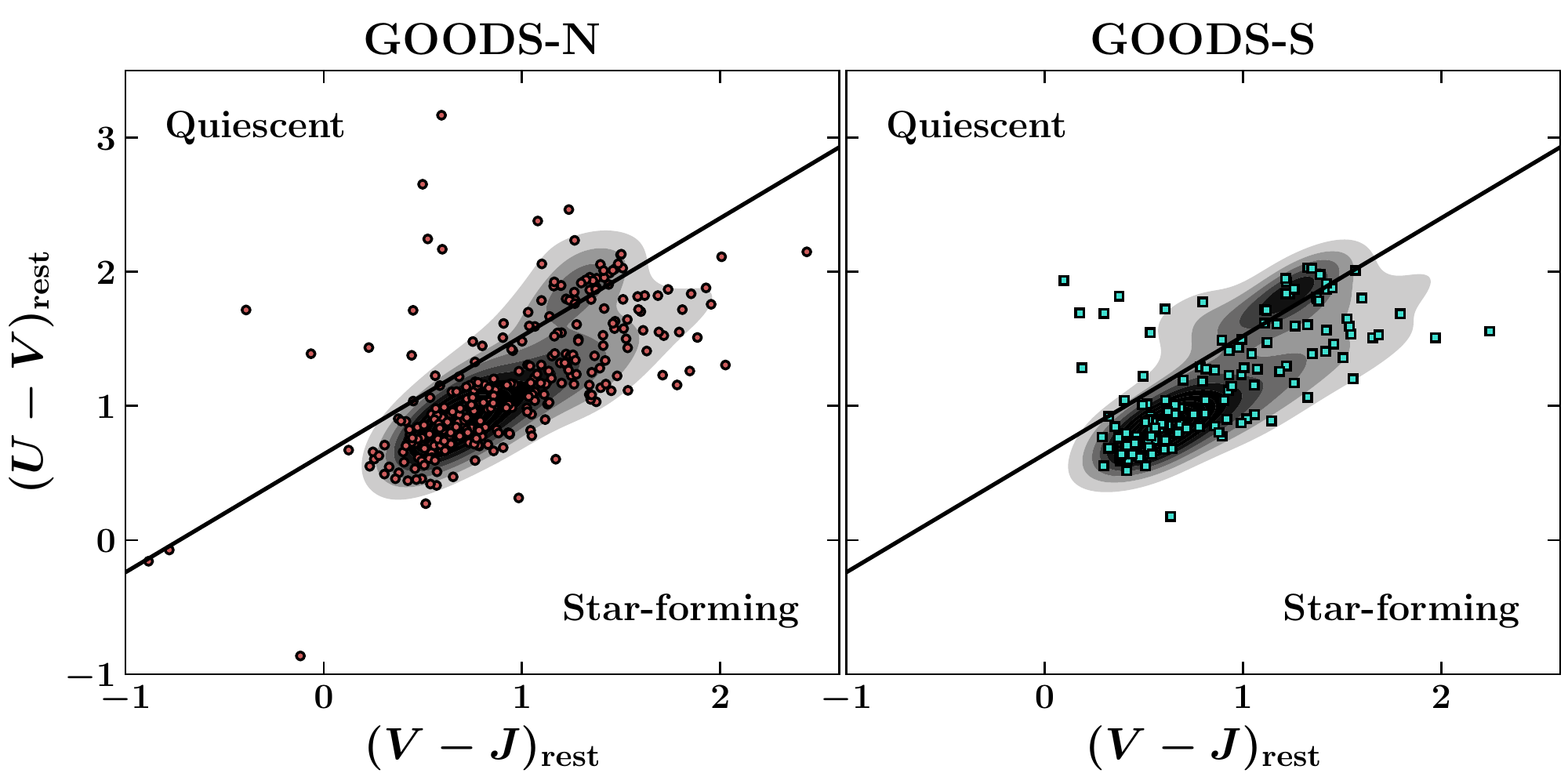}
    \caption{$UVJ$ color separation (solid black line) applied to our sample to select star-forming galaxies during the sample selection process. See Section~\ref{select_sf} for details. Background contours show the full photometrically selected sample between $0.22 \lesssim z \lesssim 0.75$ and Log$(M_{*}/\mathrm{M}_{\odot}) \geqslant 8.96$. Markers show all galaxies in CLEAR that pass the first 3 steps of sample selection (see Table~\ref{tab:sample_selection_table}) with stellar masses Log$(M_{*}/\mathrm{M}_{\odot}) \geqslant 8.96$.}
    \label{fig:uvj}
\end{figure*}

\begin{figure*}

	\centering\includegraphics[width=\textwidth]{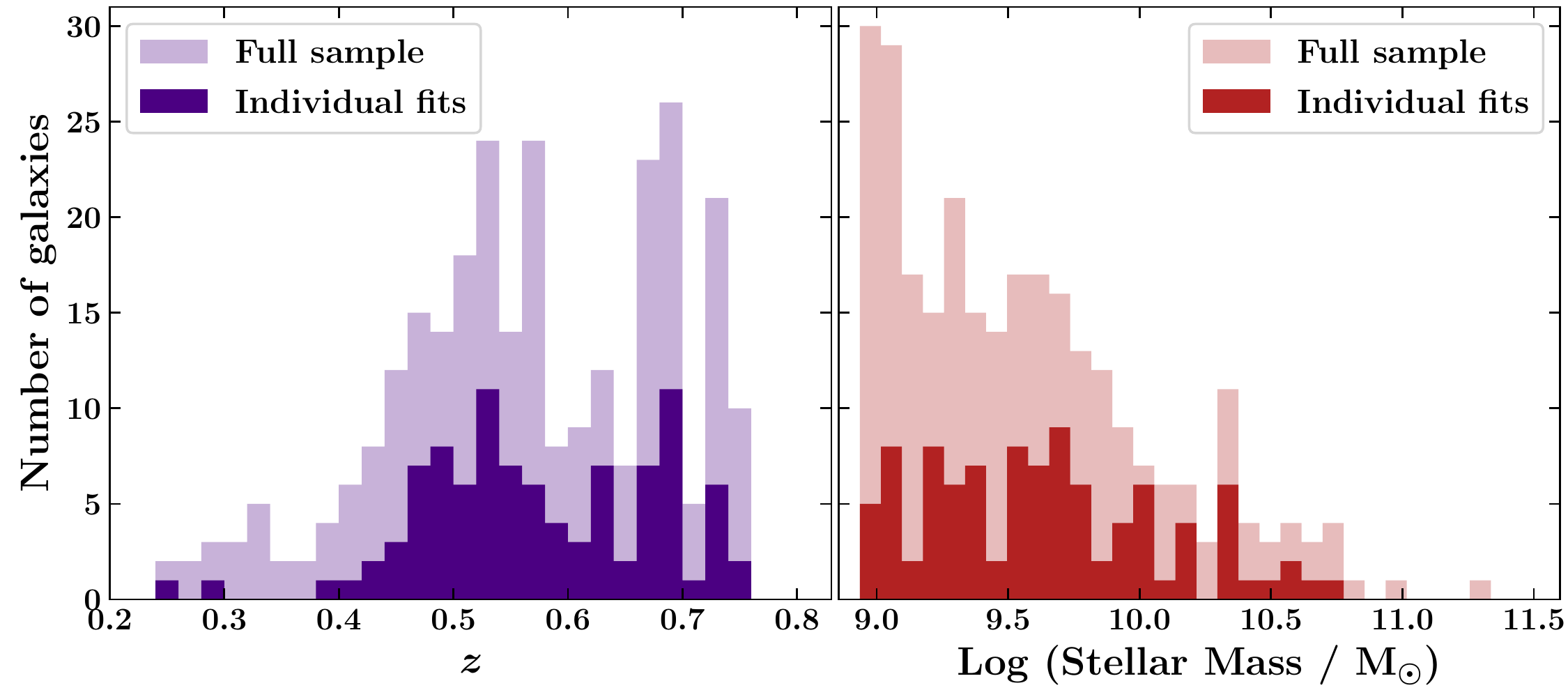}
    \caption{Grism redshift (left panel) and stellar mass (right panel) distributions for galaxies in the CLEAR sample used in this study. Full sample shows the galaxies that go into the stacks. Individual fits show the distributions for those galaxies that obtained good quality individual fits.}
    \label{fig:dists}
\end{figure*}

We then use the $UVJ$ color separation from \cite{Williams2008} to select star-forming galaxies for the appropriate redshift range of our sample. We do this by interpolating between the $0.0 < z < 0.5$ and $0.5 < z < 1.0$ $UVJ$ color separations provided in their Equation 4. The $UVJ$ color separation we use to select star-forming galaxies is stated in the fourth row of Table~\ref{tab:sample_selection_table} and is shown applied to our sample in Figure~\ref{fig:uvj}. This gives us a sample of 524 star-forming galaxies in GOODS-N and 258 in GOODS-S.

\subsubsection{Quality check}

We then quality check all 782 G102 grism spectra and the quality of the grism redshift fits by eye. This is done to ensure:

\begin{enumerate}
    \item Successful subtraction of contamination from the grism data reduction process (Section~\ref{se_cont_sub_grizli}) allowing for usable \ha~emission line maps.
    \item The grism redshift fit is not driven by poorly subtracted contamination, leading to the false identification of a \ha~line.
\end{enumerate}

After the quality check is applied, we are left with 399 star-forming galaxies in GOODS-N and 192 in GOODS-S with good quality \ha~emission line maps (a contamination rate of 24\%).

\subsubsection{Removal of Active Galactic Nuclei (AGN)}

Active Galactic Nuclei (AGN) detected from X-ray data \citep{Luo2016,Xue2016} are then removed from the sample\footnote{AGN activity can lead to \ha~emission. Since we are only interested in \ha~emission from young, hot O and B stars, we remove AGN from our sample.}. 7 are removed from GOODS-N and 2 from GOODS-S. This leaves us a sample of 392 star-forming galaxies in GOODS-N and 190 in GOODS-S. A total of 582 star-forming galaxies.

\subsubsection{Mass Completeness + H$\alpha$ Signal-to-Noise Ratio (SNR) cut}
\label{mass_snr_cut}
For our sample of 582 star-forming galaxies, we perform both individual size determination fits and stacking analysis (Section~\ref{stacking}). There is however an increasing trend in the \ha~signal-to-noise ratio (SNR) with stellar mass. Therefore, the size determination process is more challenging for low-mass galaxies.

The median stellar mass of our sample is Log$(M_{*}/\mathrm{M}_{\odot}) = 8.96$. Measurements from both individual fits and stacking analysis were the most unreliable for galaxies below this stellar mass. Therefore, the analysis in this paper focuses on star-forming galaxies with Log$(M_{*}/\mathrm{M}_{\odot}) \geqslant 8.96$ that lead to \ha~stacks of integrated SNR > 48. This amounts to a total of 279 star-forming galaxies and is referred to as the ``full sample". The distribution of this sample in grism redshifts and stellar mass can be seen in Figure~\ref{fig:dists}. We will also show the size measurements for all our reliable individual fits with Log$(M_{*}/\mathrm{M}_{\odot}) \geqslant 8.96$. The distribution for this sample is also shown in Figure~\ref{fig:dists}. It can be seen that the stellar mass distribution for the individual fits is much flatter than the stellar mass distribution for the full sample owing to the challenge of fitting low mass galaxies.

\subsubsection{All fields and GS4 Sample Separation}
\label{all_GS4_separation}

In preparation for the stacking analysis (see Section~\ref{stacking}), the full sample (Figure~\ref{fig:dists}) is separated into two groups. The first group, which we refer to as ``all fields'' contains galaxies in the full sample that have F105W imaging of similar depth. This includes all galaxies in GOODS-N and all but one CLEAR pointing in GOODS-S. The excluded GOODS-S pointing -- which we will refer to as ``GS4" -- has much deeper F105W imaging due to its overlap with the Hubble Ultra Deep Field (HUDF) and therefore many archival programs which include F105W imaging. Because we weight galaxies by their inverse variance when stacking, galaxies in GS4 completely dominate the F105W stacks. Consequently, any measurements on the F105W stacks are not representative of the full sample, but of GS4 galaxies alone which represent only 1/12 of the full sample. We therefore group GS4 galaxies separately, but conduct identical analysis on both groups.

\subsection{Stacking}
\label{stacking}

\begin{figure*}
	\centering\includegraphics[width=\textwidth]{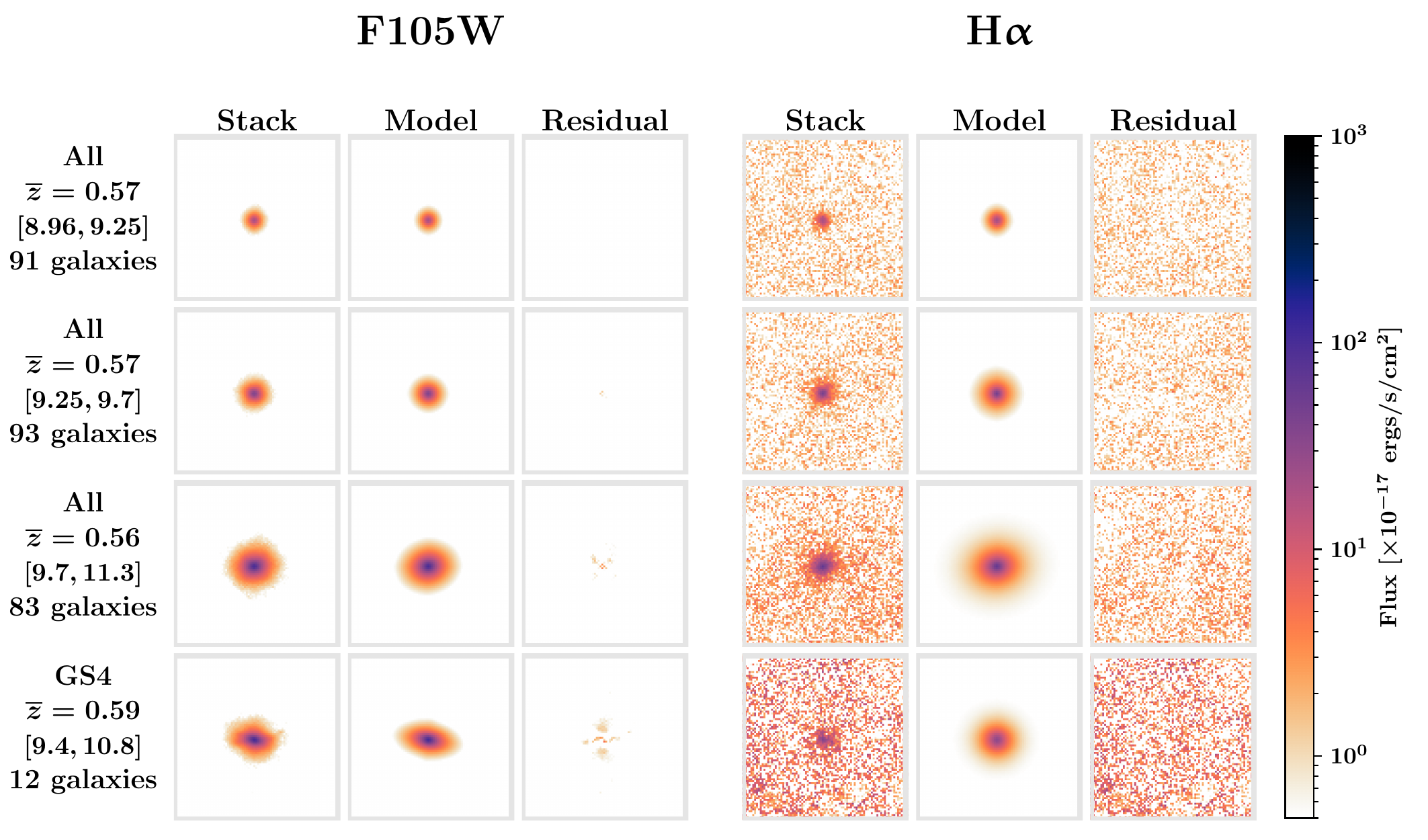}
    \caption{Zoomed-in regions ($89\times89$~pixels, pixel scale = 0.1$^{\prime\prime}$) of the stellar continuum (second column) and H$\upalpha$ (fifth column) stacks for the CLEAR sample with their associated GALFIT fits. Original thumbnails are $189\times189$~pixels (see Section~\ref{making_maps} for reasoning). The mean grism redshift ($\overline{z}$) of each stack is stated along with the Log(M$_*/\mathrm{M}_\odot)$ range of each stack in square brackets. The colormap is logarithmic, with H$\upalpha$ stacks and fits multiplied by  100 for visibility.}
    \label{fig:stacks}
\end{figure*}

Despite many of the CLEAR \ha~emission line maps having deep, 12 orbit depth observations (Section~\ref{CLEAR}) allowing for individual size measurements, the main analysis in this paper focuses on results obtained from a stacking analysis.

\subsubsection{Motivation for Stacking Analysis}
\label{just_stacking}

Reliable individual size measurements cannot be obtained for all galaxies in the full sample. This is because some of these galaxies do not lie in the maximum depth region of the CLEAR footprint given the survey design (Section~\ref{CLEAR}) and so have shallow \ha~emission line maps. Furthermore, only high SNR \ha~emission line maps would lead to reliable size measurements and these would naturally bias towards high mass galaxies due to the star formation rate -- stellar mass relation (e.g. \citealt{Whitaker2012b}). This likely leads to the \ha~SNR relation with stellar mass in the sample (Section~\ref{mass_snr_cut}), which also leads to a bias towards high mass galaxies. This introduces strong biases on conclusions made from reliable individual measurements alone and are not representative conclusions for the full sample.

Furthermore, as will be explained in Section~\ref{size_determination}, we assume a single component S\'ersic profile in our size determination process. For very deep \ha~emission line maps, this simple model can become inappropriate due to the clumpy nature of \ha~morphologies. Therefore, it leads to unreliable size measurements.

By performing a stacking analysis, all galaxies in the full sample, regardless of their observational depth, stellar mass or \ha~SNR are included. Additionally, stacking averages over the sample, smoothing out irregularities in \ha~morphologies prominent in individual maps, allowing us to focus on where most of the \ha~emission resides in typical star-forming galaxies. Thus, the \ha~morphologies of the stacks are simplified, making the single component S\'ersic profile an appropriate assumption.

Stacking also boosts SNR, allowing us to trace the \ha~distribution out to larger radii than would be possible from individual measurements.

\subsubsection{Stacking Procedure}
\label{stacking_procedure}

Our stacking method closely follows that of 3D-HST (Section~\ref{3dhst}) presented in \cite{Nelson2015}, but is more streamlined due to the sophistication of \texttt{Grizli}. Drizzled F105W direct image thumbnails and \ha~emission line maps of galaxies generated by \texttt{Grizli} in the grism data reduction process (Section~\ref{data_reduction}) are stacked in bins of stellar mass. Stellar masses for all galaxies in the CLEAR survey (Section~\ref{CLEAR}) are calculated using \texttt{Eazy-py} (EAZY, \citealt{Brammer2008}) from the wealth of ancillary multiwavelength photometry available. The ``fsps\_QSF\_12\_v3" Spectral Energy Distribution (SED) template created with \texttt{FSPS} (see Section~\ref{grism_z}) is used, assuming a \cite{Chabrier2003} Initial Mass Function (IMF). Stellar mass bins in the ``all fields'' and GS4 samples are chosen to include approximately an equal number of galaxies.

Before stacking the F105W direct image thumbnails, we convert their units from \mbox{counts s$^{-1}$} to \mbox{$\times10^{-17}$~ergs~s$^{-1}$~cm$^{-2}$}, which are the units the \ha~emission line maps are in. This is done using the pivot wavelength of the F105W filter. Bad pixels are then masked in each F105W and \ha~emission line map. Neighbouring sources to the galaxy of interest in the F105W thumbnails are masked using the \texttt{Grizli}-generated segmentation thumbnail from the \texttt{SExtractor} \citep{Bertin1996} segmentation map. The same mask is used on the corresponding \ha~emission line map. We do not use the asymmetric double pacman mask devised in \cite{Nelson2015} to mask [\ion{S}{2}] emission and poorly subtracted stellar continuum in our analysis since \texttt{Grizli} provides an improved solution. More specifically, the flux of the [\ion{S}{2}]$\lambda\lambda6717,6731$ doublet\footnote{It is assumed these lines have a 1:1 line ratio.} as well as the flux for all other emission lines are determined directly from the grism spectrum and the 2D emission line models. The models are created with the assumption that the emission line maps have the same spatial morphology as the galaxy does in F105W. Therefore the [\ion{S}{2}] emission line map is subtracted before drizzling the \ha~emission line map.

We do not correct the \ha~flux to account for the effect of blending of \ha~and [\ion{N}{2}] as was done in \cite{Nelson2015}. This is due to the fact that the majority of our galaxies have low stellar masses (Median Log$(M_{*}/\mathrm{M}_{\odot})=9.47$), where [\ion{N}{2}]/\ha~is expected to be very small -- specifically $\sim0.1$ -- at $z\sim0.5$ (e.g. \citealt{Faisst2018}). Regardless, this re-scaling reduces the overall brightness of the \ha~stack by some fixed value and so does not alter the shape of the \ha~distribution. Consequently, it does not affect structural measurements tracing morphology and size, which are the focus of our study.

\texttt{Grizli} generates inverse variance maps for both F105W and emission line map thumbnails. These provide an estimate of the errors per pixel. We use these maps to weight each pixel and then add a further weighting by F105W flux density\footnote{These are the F105W fluxes of the best-fit \texttt{Grizli} template} for each map. This second weighting ensures no single bright galaxy dominates a stack \citep{Nelson2015}. The F105W direct image thumbnails and \ha~emission line maps for each stack are summed and then exposure-corrected by dividing them by their corresponding summed weight stacks. For each stack, variance maps are therefore $\sigma_{ij}^2 = 1/\sum{w_{ij}}$, where $w_{ij}$ is the weight map for each galaxy in the stack.

Analogous work to ours as part of the 3D-HST survey (Section~\ref{3dhst}) demonstrated that rotating and aligning the thumbnails along the measured semi-major axis of the galaxy before stacking did not change their results \citep{Nelson2015}. Therefore, we do not rotate and align our thumbnails along the measured semi-major axis before stacking. 

Our resulting stacks and associated fits from the size determination process (see Section~\ref{size_determination_stacks}) can be seen in Figure~\ref{fig:stacks}. The first column entry for each stack states (in order) the sub-sample the galaxies belong to (Section~\ref{all_GS4_separation}), the mean grism redshift of the stack, the stellar mass range of galaxies in the stack and the number of galaxies in the stack. As is apparent, residuals are negligible and a clear trend of increasing size with stellar mass can be seen by eye. For the first three stacks with the largest number statistics, we can see that both the stellar continuum and \ha~emission have circular profiles. It is therefore most apparent when comparing their stacks and model fits by eye that the \ha~emission is indeed more extended than the stellar continuum. We quantitatively confirm this in Section~\ref{CLEAR_morph_stacks}.

\section{Size determination}
\label{size_determination}

As briefly mentioned in Sections~\ref{mass_snr_cut} and \ref{just_stacking}, size measurements are conducted on both individual thumbnails and stacks for both F105W and H$\upalpha$. The analysis in this paper focuses on the measurements made from the stacking analysis, since these include all the galaxies in our full sample (Figure~\ref{fig:dists}). Nevertheless, we show and discuss our size measurements for both individual thumbnails and stacks in Section~\ref{CLEAR_mass_size}. In this section, we will describe the size determination process for our individual measurements (Section~\ref{size_determination_ind}) and our stacks (Section~\ref{size_determination_stacks}).

Both size determination processes use the two-GALFIT-run approach developed and used in \cite{Matharu2018}. This approach uses \texttt{GALFIT} \citep{Peng2002b, Peng2010a} to fit 2D single-component S\'ersic profiles to the thumbnails/stacks in two iterations and leads to a high level of agreement (see Figure~\ref{fig:size_size} for level of agreement on individual F105W measurements in CLEAR and \citealt{Matharu2018}) with published size measurements for the stellar continuum from \cite{VanderWel2012}. The S\'ersic profile is defined as:

\begin{equation}
    I(r)=I(R_{\mathrm{eff}})\ \exp\left\{-\kappa\left[ \left(\frac{r}{R_{\mathrm{eff}}}\right)^{1/n}-1\right]\right\}
	\label{eq:sersic_profile}
\end{equation}
where $R_{\mathrm{eff}}$ is the effective radius. This is the radius within which half of the galaxy's total flux resides. $n$ is the S\'ersic index and $\kappa$ is an $n$-dependent parameter to ensure that half of the total intensity is contained with $R_\mathrm{eff}$. $I(R_{\mathrm{eff}})$ is the intensity at the effective radius. 

The first iteration of the two-GALFIT-run approach keeps all parameters free and obtains refined values for the shape parameters ($x$, $y$ coordinates, axis ratio and position angle) of the galaxy/stack. The second iteration fixes the values obtained for the shape parameters in the first iteration, keeping only $R_{\mathrm{eff}}$, $n$ and magnitude free. In this work, there are variations on how the two-GALFIT-run approach is implemented depending on if fitting is being conducted on direct image thumbnails or the fainter \ha~emission line maps.

\subsection{Size Determination for Individual fits}
\label{size_determination_ind}

We run our size determination process on the F105W direct image thumbnails first, since the shape parameters determined provide more reliable initial values for the \ha~emission line map size determination process.

\subsubsection{Source Detection and Initial Parameter values}
\label{source_detection_ind}

We use the \texttt{SExtractor} \citep{Bertin1996} F105W source detection results from the F105W GOODS-S and GOODS-N mosaics to obtain our initial values for F105W magnitude, axis ratio, position angle (PA) and $R_{\mathrm{eff}}$. Since \texttt{Grizli} centers all thumbnails according to their \texttt{SExtractor} determined centers in F105W, our initial $x$, $y$ coordinates for each galaxy are pixel coordinates for the center of each thumbnail. For the S\'ersic index, $n$, we use an initial value of 2.5.

\subsubsection{Point Spread Function (PSF)}
\label{psf_ind}

The resolution limit of the WFC3 leads to image smearing. A PSF image can be included in every \texttt{GALFIT} fit to account for this. We use the \texttt{Grizli} generated drizzled F105W PSFs for each galaxy. These are generated using the \cite{Anderson2016} WFC3/IR empirical PSF library. This library provides PSFs sampled on a $3\times3$ grid at various positions across the WFC3 detector. Four sub-pixel center positions are available at each of these grid points. The relevant empirical PSF is placed at the exact location of the galaxy in the detector frame. This is done for each individual exposure the galaxy is detected in. The PSF model is then drizzled to the same pixel grid as the \ha~emission line maps \citep{Mowla2018}.

\subsubsection{Noise Image}
\label{noise_ind}

To account for the noise at each pixel, a noise image -- or $\upsigma$ image in \texttt{GALFIT} nomenclature -- can be included in every \texttt{GALFIT} fit. We use the \texttt{Grizli} generated drizzled F105W and \ha~weight maps for each galaxy to create $\upsigma$ images, where $\upsigma = 1/\sqrt{\mathrm{weight}}$.

\subsubsection{Bad pixel mask}
\label{badpix_ind}

The drizzling process can sometimes lead to bad pixels with indefinite values. Not accounting for these pixels can prohibit \texttt{GALFIT} from converging to a solution other than the initial parameters provided. We account for these pixels by providing \texttt{GALFIT} with a bad pixel mask for each \texttt{GALFIT} fit.

\subsubsection{Size measurements with \texttt{GALFIT}}
\label{galfit_ind}

We run \texttt{GALFIT} on the F105W direct image thumbnails keeping all the parameters free, including fitting for the sky background. We separate the resulting fits to those that are ``successful'' and those that are ``asterisked''. Successful fits are those for which all the parameters converge to a solution. Asterisked fits are those for which there is a problem in the fitting of at least one parameter. The majority of the asterisked fits are due to problems in the fitting of the axis ratio. For this subset of problematic fits, we refit them keeping their $x$, $y$ and PA fixed, with all other parameters free. Those fits that converge in this run with no asterisks are added to the final sample of successful F105W fits.

For all the first-stage successful fits, we proceed with a second \texttt{GALFIT} run. This iteration takes the shape parameter ($x$, $y$, axis ratio and PA) values determined in the first \texttt{GALFIT} run and keeps them fixed. The parameters left free are $R_{\mathrm{eff}}$, $n$, magnitude and sky background. Those fits that converge in this run with no asterisks are added to the final sample of successful F105W fits.

All successful F105W fits are quality checked by eye, by the criteria listed below:

\begin{itemize}
    \item Over-subtraction due to the \texttt{GALFIT} model being too extended.
    \item An obvious mismatch in the PA of the \texttt{GALFIT} model to the galaxy.
    \item An obvious shape mismatch (axis ratio, axis ratio + PA) of the \texttt{GALFIT} model to the galaxy.
\end{itemize}

All three of the above criteria lead to poor residuals. However, minor problems in one or two can allow for a usable fit. No problems in all three criteria define good quality fits. Unusable fits are those where there are significant problems in all three criteria. For our sample of 582 star-forming galaxies before mass completeness and \ha~SNR cuts are applied, 497 have good quality F105W individual fits and 37 have usable F105W individual fits.

For the 534 galaxies with good quality and usable F105W individual size measurements, we proceed to make individual measurements on their corresponding \ha~emission line maps. This size determination process is more challenging due to the low SNR of \ha~emission relative to F105W. To mitigate this challenge, we make reasonable assumptions that allow us to reduce the number of parameters required to be fit for. We assume that the centroid and PA of the \ha~spatial distribution is the same as the centroid and PA of the F105W spatial distribution. Support for this assumption can be seen in the KMOS$^{\mathrm{{3D}}}$ survey (Section~\ref{kmos3d}), where only 24\% of galaxies have \ha~spatial distributions mis-aligned from their stellar continuum spatial distributions by more than $30^{\circ}$ \citep{Wisnioski2019}.

The initial \texttt{GALFIT} run on the \ha~emission line maps therefore keeps $x$, $y$ and PA fixed to values determined from the F105W size determination process. Magnitude, axis ratio, $R_{\mathrm{eff}}$ and $n$ are set to the same initial values used for the first \texttt{GALFIT} run on the F105W thumbnails (Section~\ref{source_detection_ind}). Resulting fits are separated into ``successful" and ``asterisked" categories just like in the first iteration of the F105W size determination process. \ha~emission line maps that make the successful fits category then undergo a second \texttt{GALFIT} run which fixes the axis ratio to the value determined in the first \texttt{GALFIT} run. Those fits that converge with no asterisks make the final sample of successful \ha~fits. These successful fits are then quality checked by eye using the same criteria listed above that was used on the individual F105W measurements. 164 galaxies obtained good quality \ha~individual fits and 86 obtained usable \ha~individual fits. In total, there are 152 galaxies with {\it both} good quality F105W and \ha~fits and 250 with {\it both} good quality and usable F105W and \ha~fits. Applying the mass completeness and \ha~stack SNR cuts (Section~\ref{mass_snr_cut}) to the good quality fits alone gives 97 galaxies with reliable F105W and \ha~individual fits.

Using the angular diameter distance to each galaxy calculated using its grism redshift, we convert the determined \texttt{GALFIT} $R_{\mathrm{eff}}$ from pixels to kiloparsecs. These individual measurements are shown as small red circular (F105W) and blue star (H$\upalpha$) markers in the top row of Figure~\ref{fig:ms}. These results are discussed further in Section~\ref{CLEAR_mass_size}.

\subsection{Size determination for the Stacks}
\label{size_determination_stacks}

After we complete the stacking process for the full sample (Section~\ref{stacking}), we generate the relevant files that will allow us to conduct our size determination process on the stacks directly. In this section, we describe the generation of these files and the size determination process specific to the stacking analysis.

\subsubsection{PSF construction}
\label{psf_construction}

To create PSFs for each stack, we use the \texttt{Grizli} generated drizzled F105W PSFs for each galaxy in the stack that are explained in Section~\ref{psf_ind}. A mask is created for each PSF indicating which pixels have non-zero, finite values. The PSFs for each stack are summed and then divided by the sum of the masks for that stack. The F105W PSF stack is used on both the F105W and \ha~stack for each stellar mass bin.

Note that our method for generating a PSF for each stack improves upon the method used by 3D-HST. In \cite{Nelson2015}, a single PSF modelled using the TinyTim software \citep{Krist1995} was used in the size determination process. As explained in Section~\ref{psf_ind}, the PSFs we use account for variations in the PSF across the detector frame. Stacking these PSFs for each galaxy better captures the true PSF for that particular stack.

\subsubsection{Noise estimation}
\label{noise_estimation}

During the stacking procedure (Section~\ref{stacking_procedure}), weight maps are also stacked and used to weight the pixels in each stack. These weight maps are inverse variance maps, and so can be used to generate a $\upsigma$ image for each stack. We simply take the summed weight maps for each F105W and \ha~stack, and calculate the $\upsigma$ image for the stacks, where $\upsigma_{\mathrm{stack}} = 1/\sqrt{\mathrm{weight}_{\mathrm{stack}}}$.

\subsubsection{Initial Parameter values}
\label{initial_param_stack}

As was the case for the individual fits (Section~\ref{source_detection_ind}), initial values for magnitude, axis ratio, $R_{\mathrm{eff}}$ and PA for each stack are obtained from the \texttt{SExtractor} GOODS-S and GOODS-N F105W source detection results. For each F105W stack and its corresponding \ha~stack, we use the mean value calculated from the individual values for each galaxy in F105W. The initial value for $n$ is set to 2.5 and the sky background is kept free throughout the fitting process, as was the case for the individual fits.

\subsubsection{Size measurements with \texttt{GALFIT}}
\label{galfit}

As was the case in the individual fits, we run \texttt{GALFIT} on the F105W stacks first, keeping all the parameters free (Section~\ref{galfit_ind}). The purpose of this run is to obtain refined shape parameters ($x$, $y$, axis ratio and PA), some of which can be used as initial values on the corresponding fainter, lower SNR \ha~stacks. We quality check all fits by eye. For those that fail the same quality check criteria described in Section~\ref{galfit_ind}, did not converge to a solution, or had problems with the fitting of a parameter, we rerun the fit keeping $R_{\mathrm{eff}}$ and $n$ fixed to their initial values (Section~\ref{initial_param_stack}). Fixing parameters which we are unconcerned about in the fit is a recommended strategy when \texttt{GALFIT} is struggling to converge.

After the refined shape parameters are determined from the first \texttt{GALFIT} run, we set up a second \texttt{GALFIT} run that fixes the shape parameters ($x$, $y$, axis ratio and PA) to those determined in the first \texttt{GALFIT} iteration. All these fits converge to a solution that also pass the quality check criteria described in Section~\ref{galfit_ind}. The F105W stacks and their final \texttt{GALFIT} fits can be seen in Figure~\ref{fig:stacks}.

As in the size determination process for individual fits (Section~\ref{galfit_ind}), we take the values determined for $x$, $y$, and PA for the F105W stacks and use them as initial fixed values for the corresponding \ha~stack (see Section~\ref{galfit_ind} for reasoning). All our fits converge to solutions and pass the quality check criteria outlined in Section~\ref{galfit_ind}. The second \texttt{GALFIT} run on the \ha~stacks takes the axis ratio determined in the first run and keeps it fixed, leaving $R_{\mathrm{eff}}$, $n$, magnitude and the sky background free parameters to be determined. All fits converge to sensible solutions and pass the quality check criteria described in Section~\ref{galfit_ind}. The \ha~stacks and their final \texttt{GALFIT} fits can be seen in Figure~\ref{fig:stacks}.

\section{Morphology parameters}
\label{morph_parameters}

Part of our analysis explores different morphology parameters for the stellar continuum and \ha~spatial distributions. In this section, we describe how these  parameters are calculated.

All parameters calculate the stellar mass surface density within a particular radius, assuming that the mass profile follows the light profile of the galaxy and that the light profile follows a S\'ersic profile (Equation~\ref{eq:sersic_profile}).

\subsection{The Stellar Mass Surface Density within the Effective Radius, $\Sigma_{\mathrm{eff}}$}
\label{sigma_eff}

$\Sigma_{\mathrm{eff}}$ is the stellar mass surface density within the effective radius, $R_{\mathrm{eff}}$ \citep{Barro2017a}. $R_{\mathrm{eff}}$ is defined in Section~\ref{size_determination}.

\begin{equation}
    \Sigma_{\mathrm{eff}} = \frac{0.5M_{*}}{\pi R^{2}_{\mathrm{eff}}}
\end{equation}

where $M_{*}$ is the stellar mass.

\subsection{The Stellar Mass Surface Density within 1 kpc, $\Sigma_{1\mathrm{kpc}}$}
\label{sigma_1}

$\Sigma_{1\mathrm{kpc}}$ is the stellar mass surface density within a radius of one kiloparsec \citep{Cheung2012a,Barro2017a}.

\begin{equation}
    \Sigma_{1\mathrm{kpc}} = \frac{M_{*}\gamma(2n, b_{n}R^{-1/n}_{\mathrm{eff}})}{\pi}
\end{equation}

where $M_{*}$ is the stellar mass, $n$ is the S\'ersic index. $b_{n}$ satisfies the inverse to the lower incomplete gamma function, $\gamma(2n, b_{n})$. The regularized lower incomplete gamma function $\gamma(2n, 0.5)$ is given by:

\begin{equation}
\label{eq:lower_inc_gamma}
    \gamma(2n, 0.5) = \frac{1}{\Gamma(2n)}\int_{0}^{0.5} t^{2n-1} e^{-t} dt
\end{equation}

where $\Gamma$ is the Gamma function. $\gamma(2n, b_{n}R^{-1/n}_{\mathrm{eff}})$ therefore takes the same form as Equation~\ref{eq:lower_inc_gamma}.

\subsection{The Stellar Mass Surface Density within radius $r$,  $\Sigma_{r}$}
\label{sigma_arb}

The stellar mass surface density can also be calculated within an arbitrary radius, $r$. First we define $x=b_{n}(r/R_{\mathrm{eff}})^{1/n}$ where $b_{n}$ is defined in Section~\ref{sigma_1} and $n$ is the S\'ersic index. Then the stellar mass surface density within an arbitrary radius, $\Sigma_{r}$, is:

\begin{equation}
        \Sigma_{r} = \frac{M_{*}\gamma(2n, x)}{\pi}
\end{equation}

where $\gamma$ is the regularized lower incomplete gamma function, the form of which is given in Equation~\ref{eq:lower_inc_gamma}.

\section{Results}
\label{sec:results}

In this section we discuss the results from our size measurements of the CLEAR dataset (Sections~\ref{CLEAR_mass_size} and \ref{CLEAR_morph_stacks}) and compare them to analagous results obtained from the 3D-HST and KMOS$^{\mathrm{{3D}}}$ surveys at $z\sim1$ and $z\sim1.7$ (Section~\ref{evolution}). In Section~\ref{sig1_advantages}, we discuss the advantages of using $\Sigma_{1\mathrm{kpc}}$ (Section~\ref{sigma_1}) over other morphology parameters.

\subsection{The H$\alpha$ vs. Stellar Continuum Mass--Size Relation at $z\sim0.5$}
\label{CLEAR_mass_size}

\begin{figure*}

	\centering\includegraphics[width=\textwidth]{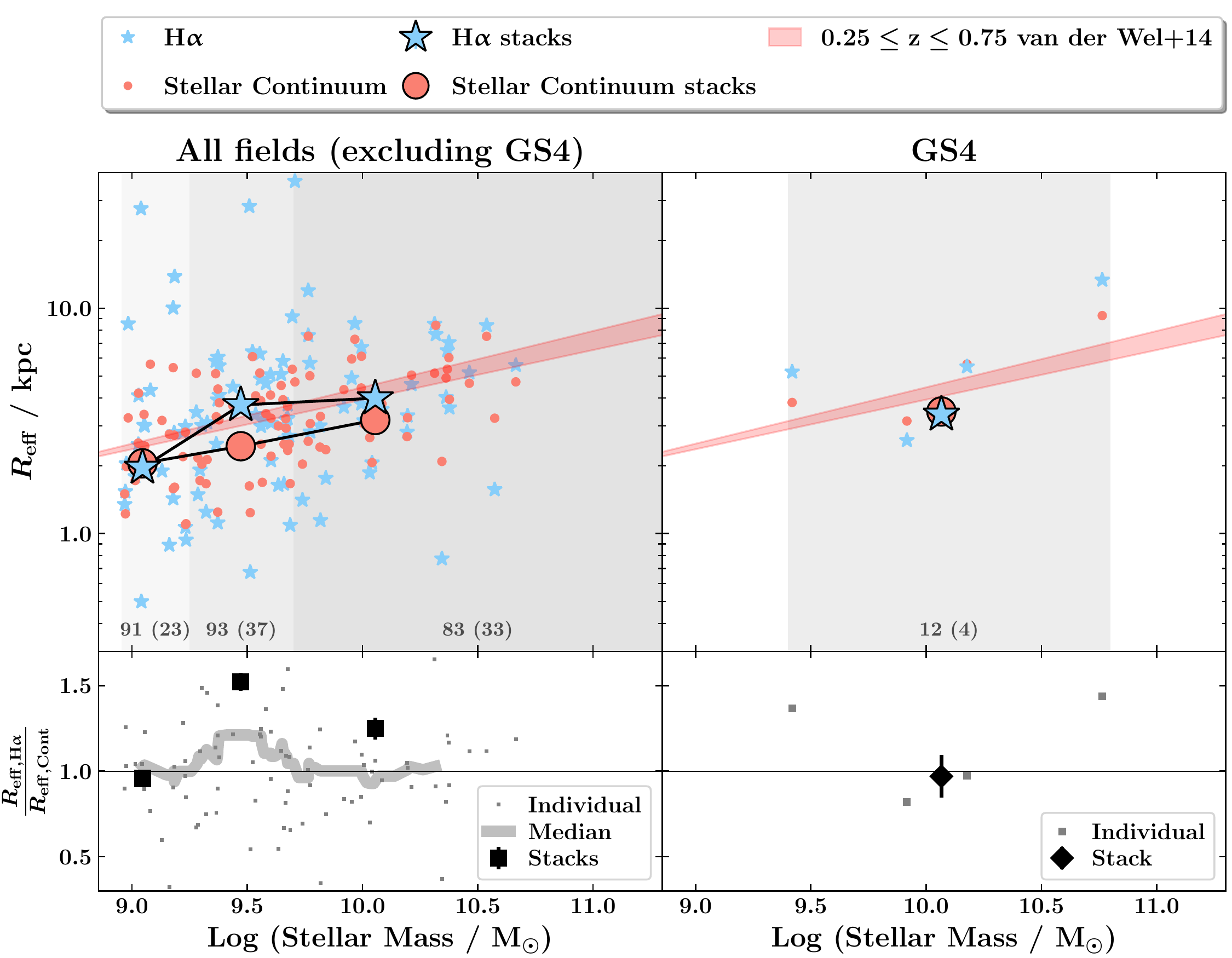}
    \caption{Top row: Stellar continuum and H$\upalpha$ stellar mass--size relations for CLEAR. Small markers show measurements on individual galaxies, larger markers show measurements from stacks. Shaded grey regions delineate the stellar mass bins for each stack. Sample sizes for the stacks and individual measurements (in brackets) are stated at the bottom for each bin. The region enclosed by the \cite{VanderWel2014} $z=0.25$ and $z=0.75$ mass--size relations (sizes measured at rest-frame wavelength of 5000{\AA}) are shown as the red shaded region. Effective radii for all measurements shown are not circularized. Bottom row: Ratio of the H$\upalpha$ to stellar continuum effective radius. The thick grey line shows a running median for the individual fits. On average, the \ha~effective radius is $1.2\pm0.1$ times larger than the effective radius of the stellar continuum at fixed stellar mass. Particularly large \ha~effective radii are seen {\it both} in the individual and stack measurements at Log$(M_{*}/\mathrm{M}_{\odot})\sim9.5$ for all fields excluding GS4. See Section~\ref{CLEAR_mass_size} for more details.}
    \label{fig:ms}
\end{figure*}

The top row of Figure~\ref{fig:ms} shows the stellar continuum (F105W) and \ha~stellar mass--size relations for CLEAR. Red circles show stellar continuum measurements and blue stars show \ha~measurements. Large markers show measurements from stacks (Section~\ref{size_determination_stacks}) whilst small markers show individual measurements (Section~\ref{size_determination_ind}). The grey shaded regions in the background delineate the stellar mass bins for each stack. The numbers at the bottom of each region state how many galaxies are in each stack, with the numbers in brackets stating the numbers of individual measurements in that stellar mass bin.

The $z=0.25$ and $z=0.75$ stellar mass--size relations for late-type galaxies at a rest-frame wavelength of 5000{\AA} from \cite{VanderWel2014} are shown as red lines with the area between them shaded red. In general, our measurements agree well with the mass--size relations derived from individual measurements in \cite{VanderWel2014} for the appropriate redshift of the CLEAR dataset. Because the measurements in \cite{VanderWel2014} are made at a rest-frame wavelength of 5000{\AA}, we show the agreement between our F105W individual measurements and those of \cite{VanderWel2012} for the same galaxies in Figure~\ref{fig:size_size} of the Appendix. No significant systematic offset between the two size determination methods exists: the mean value of our size measurements are smaller than those of \cite{VanderWel2012} by 0.4\%.
%

From the measurements on the stacks 
%
%
it can already be seen by eye that the \ha~effective radii are on average larger than those of the stellar continuum at fixed stellar mass. This is most apparent in the stacks for the ``All fields (excluding GS4)'' sample (left panel in Figure~\ref{fig:ms}).  The bottom row of Figure~\ref{fig:ms} shows this trend more explicitly in both individual and stack measurements, where the ratio of the effective radius in \ha~and the continuum is shown. Error bars on the stack measurements are calculated by jackknife re-sampling the stacks and conducting the stack size determination process (Section~\ref{size_determination_stacks}) on the jackknife re-sampled stacks.

On average, the \ha~effective radius is a factor $1.2\pm0.1$ larger than the stellar continuum effective radius for the full sample. This could be construed as evidence that 
%
%
star-forming galaxies at $z\sim0.5$ are growing in an inside-out fashion via star formation, where the short-lived O- and B-type stars tend to reside at larger galactocentric radii than the longer-lived later-type stars (we discuss this further in Section~\ref{discussion}). However, this result is driven by the Log$(M_{*}/\mathrm{M}_{\odot})\sim9.5$ stack measurement, which is higher than at any other mass albeit similar to a few measurements of individual systems. In Section~\ref{CLEAR_morph_stacks}, we explore whether this measurement is driven by peculiarities in the morphology of Log$(M_{*}/\mathrm{M}_{\odot})\sim9.5$ star-forming galaxies at $z\sim0.5$ compared to star-forming galaxies of other stellar masses.

\subsection{The Morphologies of the CLEAR H$\alpha$ and Stellar Continuum Stacks}
\label{CLEAR_morph_stacks}

\begin{figure*}

	\centering\includegraphics[width=\textwidth]{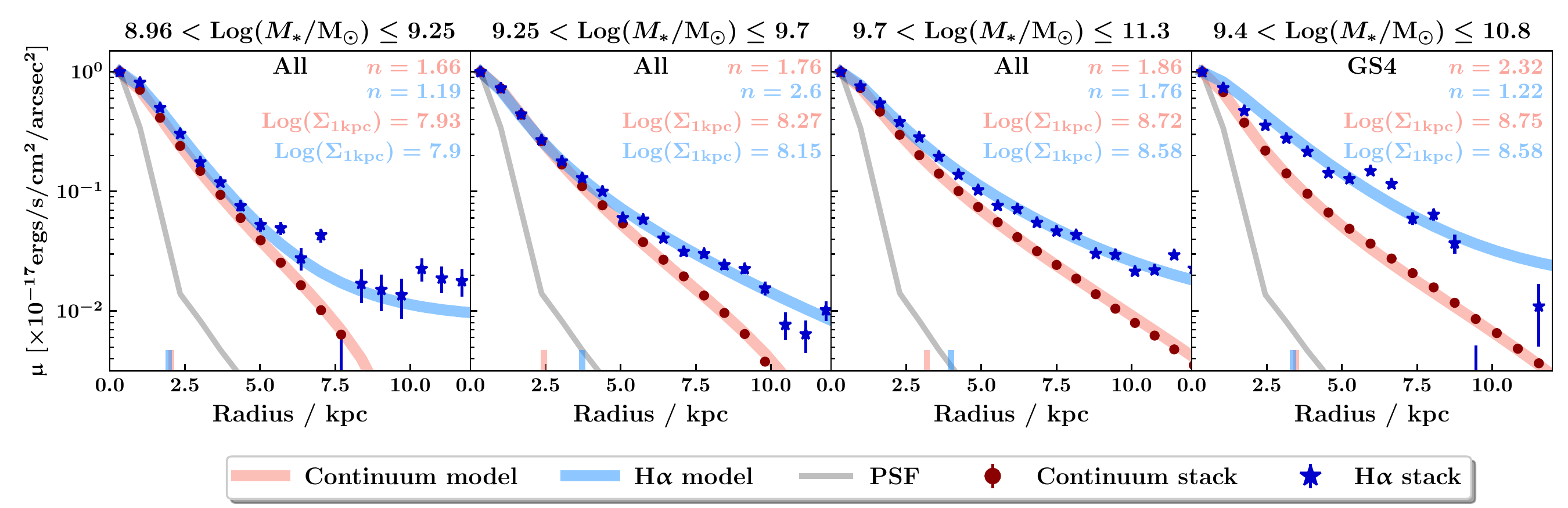}
    \caption{Normalized surface brightness profiles of the CLEAR stellar continuum and H$\upalpha$ stacks along with their GALFIT models and point spread functions (PSFs). GALFIT results for effective radii are marked with short vertical lines. GALFIT results for S\'ersic indices are stated in the top right-hand corner of each subplot. In general, GALFIT does a good job of measuring the shape of the stellar continuum and H$\upalpha$ surface brightness profiles out to $\sim10$ kpc. The stellar continuum is always more compact than the H$\upalpha$ emission when comparing the stellar mass density within 1 kpc, $\Sigma_{1\mathrm{kpc}}$. This trend breaks down when relying on S\'ersic index, $n$, as a proxy for compactness, where H$\upalpha$ seems to be more compact than the stellar continuum in one case (second panel). This reflects a degeneracy between $n$ and $R_{\mathrm{eff}}$. See Section~\ref{CLEAR_morph_stacks} for more details.}
    \label{fig:profiles}
\end{figure*}

To analyse the shape of the stellar continuum and \ha~spatial distributions for CLEAR, we measure the surface brightness profiles of the stacks and their \texttt{GALFIT} models. We use the Python package \texttt{MAGPIE}\footnote{https://github.com/knaidoo29/magpie/} to calculate all surface brightness profiles. This software has the advantages of taking into account the surface area of each pixel included within each concentric circular aperture (i.e. including fractional pixels near the edge of each aperture) and allows us to incorporate our $\upsigma$ images (Section~\ref{noise_estimation}) to appropriately calculate errors on the surface brightness profiles.

Figure~\ref{fig:profiles} shows the normalized surface brightness profiles for the stellar continuum (red circles) and \ha~(blue stars) stacks. The thick solid red and blue lines show the normalized surface brightness profiles of the associated \texttt{GALFIT} models. 2D profiles of these can be seen in Figure~\ref{fig:stacks}. The grey line shows the surface brightness profile of the PSF (Section~\ref{psf_construction}). The S\'ersic index and Log($\Sigma_{1\mathrm{kpc}}$) of each \texttt{GALFIT} model is stated in the top right-hand corner of each panel. The effective radii for both the stellar continuum and \ha~measured by \texttt{GALFIT} are marked with short vertical lines.

Out to approximately 10 kiloparsecs (1.5$^{\prime\prime}$ at $z=0.57$), the shapes of the surface brightness profiles for both the stellar continuum and \ha~stacks are captured well by \texttt{GALFIT}. The good agreement between model and data in all panels indicates that the size measurements are reliable for all mass bins. Furthermore, all surface brightness profiles are spatially resolved, since the PSF surface brightness profiles represent the resolution limit. By eye, it can be seen that the stellar continuum profiles are always steeper and therefore more compact than the \ha~profiles. This supports the more extended \ha~effective radii we measure and discuss in Section~\ref{CLEAR_mass_size}. 

S\'ersic index, $n$, is often used as a proxy for morphology (e.g. \citealt{Ravindranath2004,Bell2008a,Blanton2009,Bell2012,Papovich2015}). However, it is somewhat unreliable due to its degeneracy with the effective radius. This degeneracy means S\'ersic index and effective radius are highly covariant (e.g. \citealt{Ji2020,Ji2021}). We see an example of this issue in the second panel of Figure~\ref{fig:profiles}. For all other stellar mass bins, the S\'ersic index for the \ha~profiles is smaller than for the stellar continuum profiles ($n(\mathrm{H}\upalpha)<n(\mathrm{Cont})$), supporting the case for more extended \ha~reflected by the larger \ha~effective radii measurements discussed in Section~\ref{CLEAR_mass_size}.
This trend breaks down for the stellar mass bin within which we see particularly large \ha~effective radii measurements. $n(\mathrm{H}\upalpha)>n(\mathrm{Cont})$, even though we can see by eye that overall the \ha~surface brightness profile is more extended than the stellar continuum surface brightness profile.

One way to remove the degeneracy between $n$ and $R_{\mathrm{eff}}$ is to combine them into one morphology parameter. An example of such a parameter is the stellar mass surface density within 1 kiloparsec, $\Sigma_{1\mathrm{kpc}}$, described in Section~\ref{sigma_1} \citep{Cheung2012a,Barro2017a,Estrada-Carpenter2020}. When $\Sigma_{1\mathrm{kpc}}$ is calculated at a particular wavelength, it indicates what the stellar mass surface density within 1 kpc would be if the stellar mass were traced by the flux at that wavelength. Comparing $\Sigma_{1\mathrm{kpc}}$ calculated at two different wavelengths is equivalent to comparing the fraction of total light emitted within 1 kpc at those two wavelengths. We calculate $\Sigma_{1\mathrm{kpc}}$ for all our stacks and state the values for Log$(\Sigma_{1\mathrm{kpc}})$ under those of the S\'ersic indices in each panel of Figure~\ref{fig:profiles}. Larger values of $\Sigma_{1\mathrm{kpc}}$ indicate a higher degree of compactness within a radius of 1 kpc. We can see that the \ha~and stellar continuum trends in Log$(\Sigma_{1\mathrm{kpc}})$ better reflect the relative shape of the \ha~and stellar continuum surface brightness profiles. In every case, the stellar continuum is more compact than \ha~within 1 kpc. This result quantitatively confirms the initial impression from Figure~\ref{fig:stacks} that the \ha~emission is more extended than the stellar continuum. In Section~\ref{sig1_advantages}, we discuss the advantages of $\Sigma_{1\mathrm{kpc}}$ as a morphology parameter and advocate for its use in similar future studies.

In the next Section, we will compare our measurements with CLEAR at $z\sim0.5$ to those of 3D-HST at $z\sim1$ and KMOS$^{\mathrm{{3D}}}$ at $z\sim1.7$ to study the evolution of spatially resolved star formation in star-forming galaxies over a wide range in redshift.

\subsection{The Evolution of Spatially Resolved Star Formation in Star-forming Galaxies between $0.5\lesssim z \lesssim1.7$}
\label{evolution}

\begin{figure}
	\centering\includegraphics[width=\columnwidth]{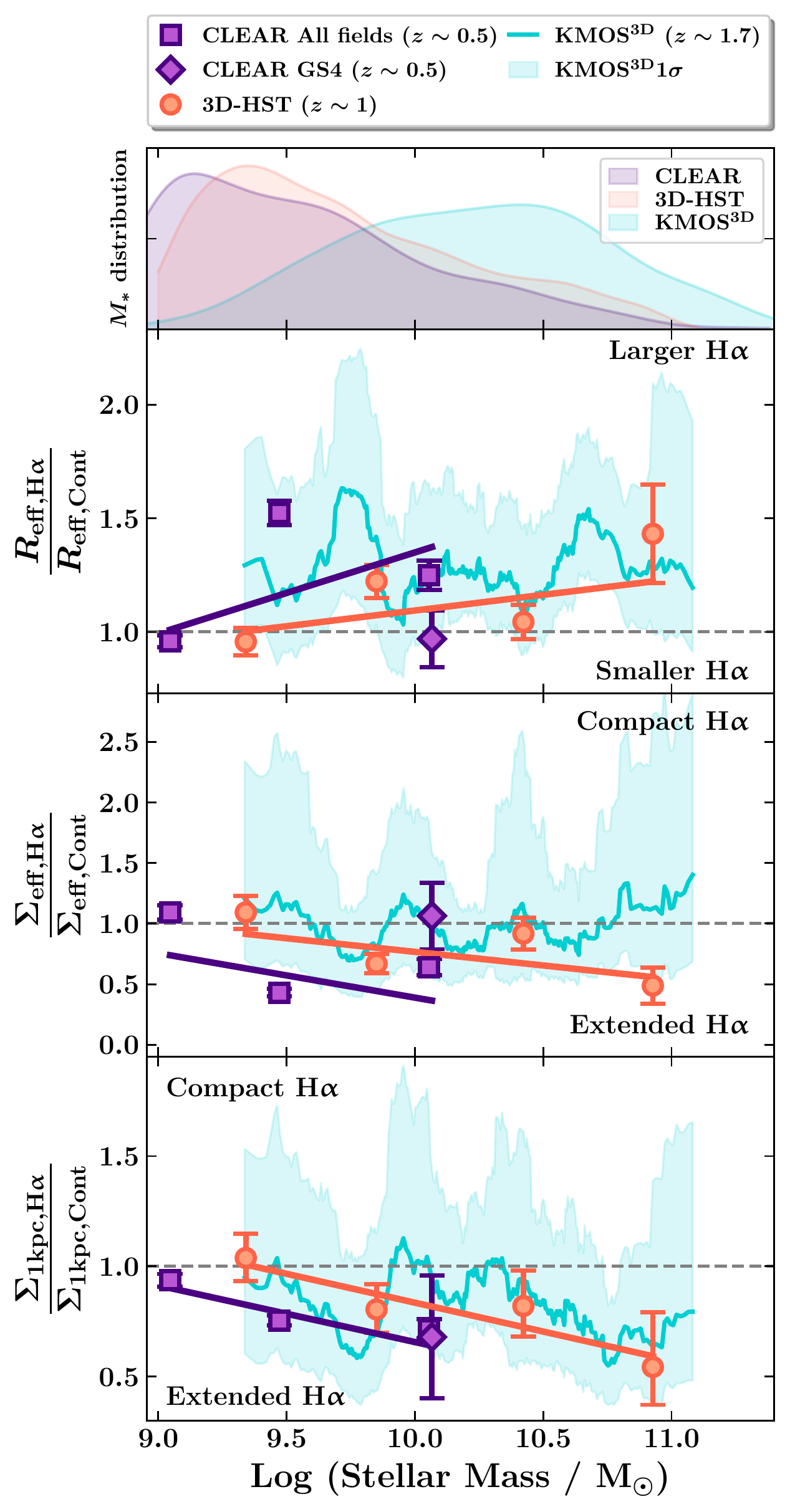}
    \caption{Evolution of H$\upalpha$ to stellar continuum morphologies for star-forming galaxies between $0.5\lesssim z \lesssim1.7$. Top panel: Stellar mass distributions for the different datasets. Ratio of the H$\upalpha$ to stellar continuum effective radius $R_{\mathrm{eff}}$ (second panel), surface density within the effective radius $\Sigma_{\mathrm{eff}}$ (third panel) and surface density within 1 kiloparsec $\Sigma_{1\mathrm{kpc}}$ (bottom panel) for CLEAR ($z\sim0.5$), 3D-HST ($z\sim1$) and KMOS$^{3\mathrm{D}}$ ($z\sim1.7$) star-forming galaxies. KMOS$^{3\mathrm{D}}$ results are plotted as running means from individual measurements. Over the stellar mass range shown, the H$\upalpha$ to stellar continuum $R_{\mathrm{eff}}$, $\Sigma_{\mathrm{eff}}$ and $\Sigma_{1\mathrm{kpc}}$ are on average the same between $0.5\lesssim z\lesssim 1.7$ within $1\sigma$ (see Table~\ref{tab:means}). Trends of larger/more extended H$\upalpha$ than stellar continuum for higher mass galaxies at fixed redshift and larger/more extended H$\upalpha$ for $z\sim0.5$ galaxies at fixed stellar mass are apparent for all three morphology parameters, but are stronger in $\Sigma_{1\mathrm{kpc}}$. Between $0.5\lesssim z\lesssim 1$, the slope of the $\frac{\Sigma_{1\mathrm{kpc, H}\upalpha}}{\Sigma_{1\mathrm{kpc, Cont}}}$--stellar mass relation is constant, but the intercept increases by a factor of 1.07 (see Table~\ref{tab:lines}) between $z\sim0.5$ and $z\sim1$ (bottom panel). At Log$(M_{*}/\mathrm{M}_{\odot})=9.5$, star-forming galaxies at $z\sim0.5$ have a $(19\pm2)\%$ lower surface density in \ha~compared to the stellar continuum than $z\sim1$ star-forming galaxies within a galactocentric radius of 1 kiloparsec.}
    \label{fig:comp}
\end{figure}

\begin{deluxetable}{cccc}
\tablecaption{Mean H$\upalpha$/Continuum size and morphology ratios. $R_{\mathrm{eff}}$, $\Sigma_{\mathrm{eff}}$ and $\Sigma_{1\mathrm{kpc}}$ are the effective radius, the stellar mass surface density within the effective radius (Section~\ref{sigma_eff}) and the stellar mass surface density within 1 kiloparsec (Section~\ref{sigma_1}). \label{tab:means}}
\tablewidth{0pt}
\tablehead{
\colhead{H$\upalpha$/Continuum ratio} & \colhead{CLEAR} & \colhead{3D-HST} & \colhead{KMOS$^{3\mathrm{D}}$} \\
\colhead{} & \colhead{$z\sim0.5$} & \colhead{$z\sim1$} & \colhead{$z\sim1.7$}
}
\startdata
$R_{\mathrm{eff, H}\upalpha}$/$R_{\mathrm{eff, Cont}}$ & $1.18\pm0.08$ & $1.16\pm0.12$ & $1.29_{-0.39}^{+0.92}$  \\
$\Sigma_{\mathrm{eff, H}\upalpha}$/$\Sigma_{\mathrm{eff, Cont}}$ & $0.81\pm0.15$ & $0.79\pm0.13$ & $1.00_{-0.81}^{+2.89}$  \\
$\Sigma_{1\mathrm{kpc, H}\upalpha}$/$\Sigma_{1\mathrm{kpc, Cont}}$ & $0.77\pm0.14$ & $0.80\pm0.17$ & $0.84_{-0.69}^{+1.80}$  \\
\enddata
\end{deluxetable}

\begin{deluxetable*}{lcccccccc}
\tablecaption{Least-Squares linear fits to the $R_{\mathrm{eff, H}\upalpha}$/$R_{\mathrm{eff, Cont}}$--, $\Sigma_{\mathrm{eff, H}\upalpha}$/$\Sigma_{\mathrm{eff, Cont}}$-- and $\Sigma_{1\mathrm{kpc, H}\upalpha}$/$\Sigma_{1\mathrm{kpc, Cont}}$--stellar mass distributions.
\label{tab:lines}}
\tablewidth{0pt}
\tablehead{
\colhead{Dataset} & \colhead{$z$} & \colhead{Log$(M_{*}/\mathrm{M}_{\odot})$} & \multicolumn2c{$R_{\mathrm{eff, H}\upalpha}$/$R_{\mathrm{eff, Cont}}$} & \multicolumn2c{$\Sigma_{\mathrm{eff, H}\upalpha}$/$\Sigma_{\mathrm{eff, Cont}}$} & \multicolumn2c{$\Sigma_{1\mathrm{kpc, H}\upalpha}$/$\Sigma_{1\mathrm{kpc, Cont}}$} \\
\colhead{} & \colhead{} & \colhead{Median} & \colhead{Gradient} & \colhead{Intercept} & \colhead{Gradient} & \colhead{Intercept} & \colhead{Gradient} & \colhead{Intercept}
}
\startdata
CLEAR & $0.22 \lesssim z \lesssim 0.75$ & 9.47 & $0.36\pm0.12$ & $-2.22\pm10.1$ & $-0.37\pm0.31$ & $4.05\pm28.3$  & $-0.26\pm0.01$ & $3.22\pm0.83$ \\
3D-HST & $0.7<z<1.5$ & 9.62 & $0.14\pm0.03$ & $-0.29\pm2.45$ & $-0.22\pm0.05$ & $3.01\pm5.42$  & $-0.26\pm0.01$ & $3.45\pm0.77$ \\
\enddata
\end{deluxetable*}

Including CLEAR, there are now a few studies on spatially resolved star formation of $z>0$ star-forming galaxies using \ha~emission line maps \citep{Nelson2015,Tacchella2015a,Wilman2020}. The redshift ranges of these studies make it possible to study the evolution of spatially resolved star formation in star-forming galaxies over a wide range in redshift for the first time.

Figure~\ref{fig:comp} compares the results from CLEAR ($z\sim0.5$, this work) to the results of 3D-HST ($z\sim1$, \citealt{Nelson2015}) and KMOS$^{\mathrm{{3D}}}$ ($z\sim1.7$, \citealt{Wilman2020}). For each of these studies, we compare the effective radii ($R_{\mathrm{eff}}$), stellar mass surface densities within the effective radius ($\Sigma_{\mathrm{eff}}$,  Section~\ref{sigma_eff}) and within 1~kpc ($\Sigma_{1\mathrm{kpc}}$, Section~\ref{sigma_1}) derived for the \ha~emission line maps and the stellar continuum.

It is worth comparing the properties of the galaxy samples and data quality between CLEAR, 3D-HST, and KMOS$^\mathrm{3D}$.  CLEAR (Section~\ref{CLEAR}) and 3D-HST (Section~\ref{3dhst}) are the most similar in terms of the type of dataset and methodology. Both use data from the same telescope (HST) and instrument (WFC3) and use the same technique of grism spectroscopy. The only difference is the wavelength range and depth of the imaging (F105W vs F140W) and spectroscopy (G102 vs G141). The CLEAR work follows the same methodology as 3D-HST, but deviates in places where \texttt{Grizli} provides improved solutions. The stellar mass distribution of the galaxies in both samples is similar (Top panel of Figure~\ref{fig:comp}). 

KMOS$^{\mathrm{{3D}}}$ (Section~\ref{kmos3d}) on the other hand uses \ha~emission line maps obtained from ground-based IFU observations with KMOS on the VLT. These have coarser spatial resolution (see Section~\ref{kmos3d}) compared to emission line maps obtained from HST WFC3 slitless spectroscopy. \texttt{GALFIT} is not used for the size determination process and it is assumed that the \ha~emission follows a pure exponential profile ($n=1$). Nevertheless, \cite{Wilman2020} demonstrate the high level of agreement between their size determination method and that of \cite{VanderWel2014} for the same set of galaxies in their stellar continuum (HST WFC3 F160W). Furthermore, CLEAR and 3D-HST use the stacking technique in bins of stellar mass where as the KMOS$^{\mathrm{{3D}}}$ measurements are made on individual galaxies. The stellar mass distribution of the KMOS$^{\mathrm{{3D}}}$ sample is the most different to that of CLEAR and 3D-HST, as it is skewed to higher stellar masses (Top panel of Figure~\ref{fig:comp}). The median stellar masses in Log$(M_{*}/\mathrm{M}_{\odot})$ for the CLEAR, 3D-HST and KMOS$^{\mathrm{{3D}}}$ samples are 9.47, 9.62 and 10.22 respectively.

\subsubsection{Results using the Effective Radius, $R_{\mathrm{eff}}$}
\label{reff_results}

In the second panel of Figure~\ref{fig:comp}, we show the ratio of the \ha~$R_{\mathrm{eff}}$ to that of the stellar continuum $R_{\mathrm{eff}}$ for all three datasets. We see that especially for 3D-HST, there is a trend that the \ha~sizes are larger than the stellar continuum sizes for galaxies at higher mass. This trend is apparent, but weaker for CLEAR. More quantitatively, the linear fit to the CLEAR measurements also indicates a positive trend but with a lower level of significance than the linear fit to the 3D-HST measurements. This can be seen explicitly in Table~\ref{tab:lines}. The KMOS$^{\mathrm{{3D}}}$ measurements however do not exhibit this trend. On average, all three data sets agree within $1\upsigma$ of each other. This is shown explicitly in the first row of Table~\ref{tab:means}, where we state the mean $R_{\mathrm{eff, H}\upalpha}$/$R_{\mathrm{eff, Cont}}$ for each data set and their errors. 

\subsubsection{Results using the Surface Density within the Effective Radius, $\Sigma_{\mathrm{eff}}$}
\label{sigeff_results}

The third panel of Figure~\ref{fig:comp} shows the ratio of the \ha~$\Sigma_{\mathrm{eff}}$ (Section~\ref{sigma_eff}) to that of the stellar continuum $\Sigma_{\mathrm{eff}}$ for all three datasets. As was the case for the trends in $R_{\mathrm{eff}}$ (Section~\ref{reff_results}), high mass galaxies are more extended in \ha~than their stellar continuum in both CLEAR and 3D-HST. The trend is stronger over the stellar mass range of 3D-HST. Once again we provide the linear fits shown for both datasets in Table~\ref{tab:lines}. The KMOS$^{\mathrm{{3D}}}$ measurements however do not exhibit a significant trend, having a $\Sigma_{\mathrm{eff, H}\upalpha}$/$\Sigma_{\mathrm{eff, Cont}}$ that is consistent with unity over the stellar mass range probed (see also Table~\ref{tab:means}), especially in the region where the stellar mass distribution is well sampled.

\subsubsection{Results using the Surface Density within 1~kpc, $\Sigma_{1\mathrm{kpc}}$}
\label{sig1_results}

In Section~\ref{CLEAR_morph_stacks}, we explained how the effective radius is degenerate with the S\'ersic index. This contributes to the scatter between stellar mass versus the effective radius and $\Sigma_{\mathrm{eff}}$ in the second and third panels of Figure~\ref{fig:comp}. $\Sigma_{1\mathrm{kpc}}$ however does not suffer from this degeneracy because it combines the two degenerate morphology parameters into a single one (see Section~\ref{sigma_1}). We therefore present our measurements for $\Sigma_{1\mathrm{kpc}}$ for all three datasets and compare them to our $R_{\mathrm{eff}}$ and $\Sigma_{\mathrm{eff}}$ measurements in Figure~\ref{fig:comp}.


In the bottom panel of Figure~\ref{fig:comp}, we show the ratio of the \ha~$\Sigma_{1\mathrm{kpc}}$ to that of the stellar continuum $\Sigma_{1\mathrm{kpc}}$ for all three datasets (CLEAR, 3D-HST, and KMOS$^\mathrm{3D}$).  Using this morphology parameter we see a clear (pun not intended) negative trend in all three datasets. More massive galaxies at all redshifts have lower \ha~surface densities compared to their stellar continuum. We discuss this result along with its physical implications in more detail in Section~\ref{discussion}. More importantly, the linear trends found for both the CLEAR and 3D-HST measurements are much tighter than those calculated for $R_{\mathrm{eff}}$ and $\Sigma_{\mathrm{eff}}$. This can be seen in the high significance of the gradient and intercepts determined in Table~\ref{tab:lines}. The high significance of the $\Sigma_{1\mathrm{kpc, H}\upalpha}$/$\Sigma_{1\mathrm{kpc, Cont}}$--stellar mass relations at $z\sim0.5$ and $z\sim1$ as well as the apparent non-evolving slope unveil a clear (pun intended) redshift trend at fixed stellar mass. Star-forming galaxies at $z\sim0.5$ have lower surface densities in \ha~emission compared to the surface densities in their stellar continuum than their $z\sim1$ counterparts at fixed stellar mass within a galactocentric radius of 1 kpc. More specifically, at a fixed stellar mass of Log$(M_{*}/\mathrm{M}_{\odot})=9.5$, star-forming galaxies at $z\sim0.5$ have a $(19\pm2)\%$ lower surface density in \ha~emission compared to the stellar continuum surface density within 1~kpc than $z\sim1$ star-forming galaxies.

In the next Section, we discuss why these trends likely became apparent when using the $\Sigma_{1\mathrm{kpc}}$ morphology parameter and discuss further advantages of its use in such studies.

\subsection{The Advantages of $\Sigma_{1\mathrm{kpc}}$ as a Morphology Parameter}
\label{sig1_advantages}

\begin{figure*}
	\centering\includegraphics[width=\textwidth]{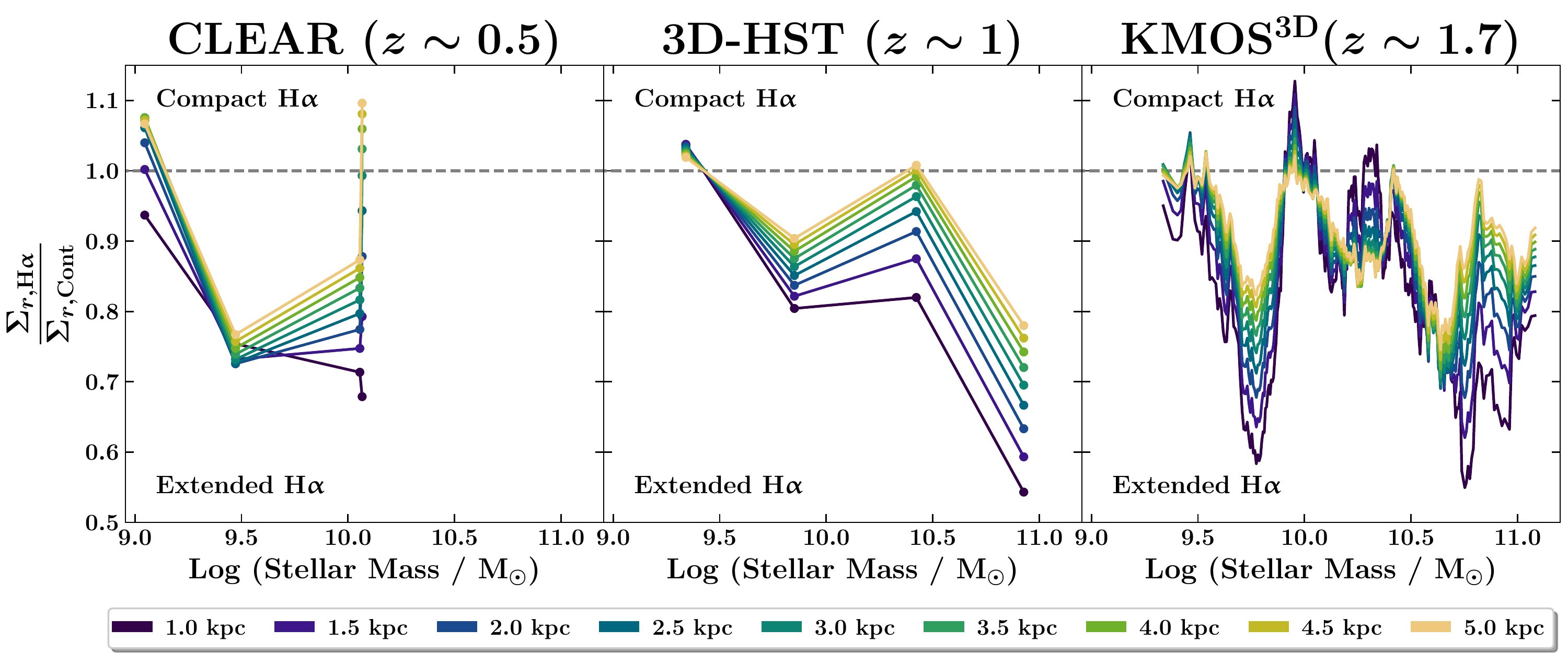}
    \caption{The ratio of the H$\upalpha$ to stellar continuum stellar mass surface density within various galactocentric radii, $\Sigma_{r}$. $r=5~\mathrm{kpc}$ represents the approximate maximum effective radius measured in the three datasets. $\Sigma_{1\mathrm{kpc}}$ (dark purple lines) exhibits the strongest trend with stellar mass for all three datasets. $\Sigma_{r}$ at smaller galactocentric radii are more reliable since they probe the highest signal-to-noise-ratio and therefore most well-determined parts of the stellar continuum and H$\upalpha$ light profiles (see Figure~\ref{fig:profiles} and Section~\ref{sig1_advantages} for more details).}
    \label{fig:sigma_all}
\end{figure*}

In general, the stellar mass surface density within some arbitrary radius, $\Sigma_{r}$ --- whether that radius be one kiloparsec or not --- is a more reliable morphology parameter to use than $n$, $R_{\mathrm{eff}}$ or even $\Sigma_{\mathrm{eff}}$. This is because it breaks the parameter degeneracy between $n$ and $R_{\mathrm{eff}}$ discussed in Section~\ref{CLEAR_morph_stacks}. $n$, $R_{\mathrm{eff}}$ and any parameter that is calculated using either $n$ or $R_{\mathrm{eff}}$ (such as $\Sigma_{\mathrm{eff}}$), is effected by this degeneracy. This parameter degeneracy increases scatter in $n$ and $R_{\mathrm{eff}}$ measurements, rendering some measurements unreliable. An example of this issue can be seen in Figure~\ref{fig:comp}, where the particularly large \ha~$R_{\mathrm{eff}}$ measurements at Log$(M_{*}/\mathrm{M}_{\odot})\sim9.5$ in CLEAR (Figure~\ref{fig:ms}) can be seen in both $R_{\mathrm{eff}}$ and $\Sigma_{\mathrm{eff}}$ measurements, but not in $\Sigma_{1\mathrm{kpc}}$ measurements. It can also be seen that the level of scatter at fixed stellar mass in $\Sigma_{1\mathrm{kpc}}$ is far lower than in $R_{\mathrm{eff}}$ and $\Sigma_{\mathrm{eff}}$ (see also \citealt{Cheung2012a,Fang2013,VanDokkum2014,Tacchella2015,Woo2015a,Whitaker2017,Barro2017a}). The least-squares linear fits for CLEAR and 3D-HST go through all the points and more quantitatively, the derived gradient and intercept values have higher significance (Table~\ref{tab:lines}) than the trends measured from $R_{\mathrm{eff}}$ and $\Sigma_{\mathrm{eff}}$ measurements.

Because $\Sigma_{r}$ mitigates the $n-R_{\mathrm{eff}}$ parameter degeneracy, there is no apparent reason as to why $r=1~\mathrm{kpc}$ should be preferred to any other radius. In Figure~\ref{fig:sigma_all}, we explore this using $\Sigma_{r}$ measured at different radii, $r=1.0$ to 5.0 kpc (where $5$~kpc is approximately the maximum value of the effective radii measured for the galaxies in all three datasets). The strongest trends in $\Sigma_{r}$ with stellar mass are apparent for lower values of $r$, with $\Sigma_{1\mathrm{kpc}}$ exhibiting the strongest trend.

Interestingly, in all three datasets, $\Sigma_{r}$ behaves somewhat differently at Log$(M_{*}/\mathrm{M}_{\odot})=9.5$. $\Sigma_{1\mathrm{kpc}}$ has the highest or higher \ha~surface density compared to $\Sigma_{r>1\mathrm{kpc}}$. This may be indicative of central starburst activity, which is prevalent for star-forming galaxies toward low stellar masses \citep{Fumagalli2012,Khostovan2021}. $\Sigma_{r}$ towards larger $r$ asymptotes towards the trends we see in $\Sigma_{\mathrm{eff}}$ (third panel, Figure~\ref{fig:comp}). There are a few reasons that can explain this, which we discuss in the sections that follow.

\subsubsection{Unreliability of $\Sigma_{r}$ at large radii}
\label{unreliable_larger}
Regarding $\Sigma_r$, there are two effects that act at larger radii and render $\Sigma_r$ less reliable for studies like that in this work.
%
The first is that the stellar continuum and \ha~light profiles are more uncertain at larger radii. This can be seen in Figure~\ref{fig:profiles}, where the error bars on surface brightness of the \ha~emission in the stacked sample increase substantially at larger radii.
%
As we calculate $\Sigma_{r}$ for $r$ at larger radii,  we are increasingly including low SNR regions of the light profile, introducing more uncertainty in our $\Sigma_{r}$ measurement. Any $\Sigma_{r}$ trends with stellar mass become increasingly washed out towards larger radii. \cite{Tilvi2013} showed that the SNR within an aperture reaches a maximum at approximately 0.69$\times$FWHM. The CLEAR, 3D-HST and KMOS$^{3\mathrm{D}}$ data have spatial resolutions with FWHMs $\sim$ $0.128^{\prime\prime}$, $0.141^{\prime\prime}$ and $0.456^{\prime\prime}$ respectively (see Sections~\ref{CLEAR}, \ref{3dhst} \& \ref{kmos3d}). The aperture within which the SNR is maximum for sources in these surveys are then $0.09^{\prime\prime}$, $0.1^{\prime\prime}$ and $0.3^{\prime\prime}$. At the respective redshifts of these surveys, this is equal to 0.6~kpc, 0.8~kpc and 2.6~kpc. It is therefore unsurprising that $\Sigma_{1\mathrm{kpc}}$ which encompasses or is measured well within these radii exhibits the strongest trends with stellar mass in Figure~\ref{fig:sigma_all}.

The second effect is that the ratio of the \ha~to stellar continuum $\Sigma_{r}$, $\Sigma_{r\mathrm{, H}\upalpha}$/$\Sigma_{r\mathrm{, Cont}}$ asymptotes to unity at large radii, since both light profiles are normalized to the total flux in their respective wavelength ranges
when calculating $\Sigma_{r}$ (see Section~\ref{morph_parameters}). $\Sigma_{r\mathrm{, H}\upalpha}$/$\Sigma_{r\mathrm{, Cont}}$ at large radii is therefore less sensitive to differences between the shapes of the stellar continuum and \ha~light profiles.

Both effects discussed above advocate for choosing a smaller value of $r$ when using $\Sigma_{r}$.  This improves the ability of this quantity to reveal trends that are otherwise washed out at large radii.

\section{Discussion}
\label{discussion}

\begin{figure*}
	
	\centering\includegraphics[width=\textwidth]{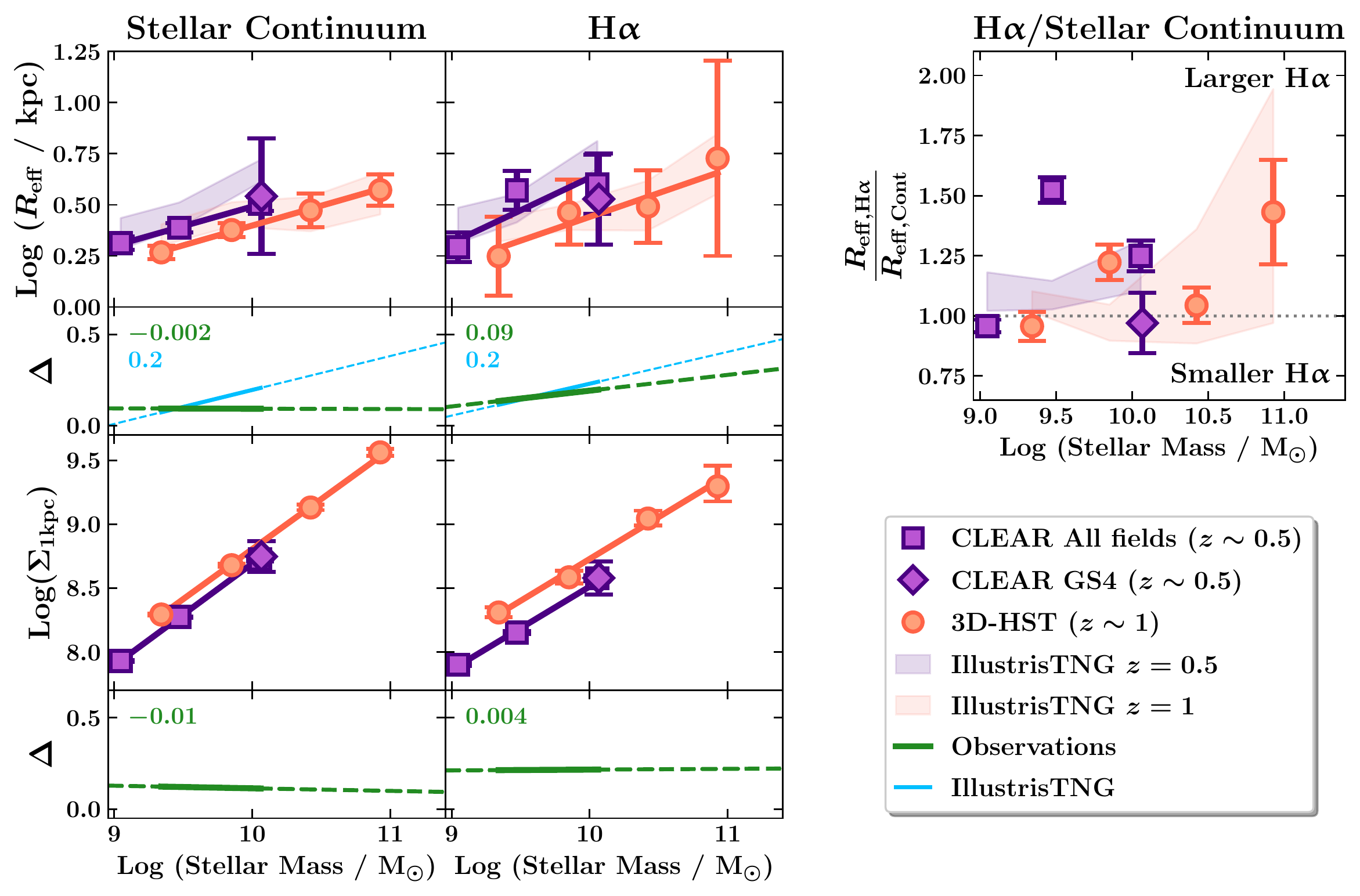}
	\caption{Left: The stellar mass--size (top row) and $\Sigma_{1\mathrm{kpc}}$--stellar mass relations (third row) for the stellar continuum and \ha~emission measured at $z\sim0.5$ from CLEAR and at $z\sim1$ from 3D-HST. Purple and orange shaded regions show $z=0.5$ and $z=1$ TNG50 measurements from \cite{Pillepich2019} applied to a galaxy sample matched to the observed ones. Second and fourth rows show Log($R_{\mathrm{eff, CLEAR}}$)-Log($R_{\mathrm{eff, 3D-HST}}$) and Log($\mathrm{\Sigma}_{1\mathrm{kpc, 3D-HST}}$)-Log($\mathrm{\Sigma}_{1\mathrm{kpc, CLEAR}}$) respectively, with the solid line showing the stellar mass range over which there are measurements from both surveys/comparable samples from TNG50. The numbers are the gradients of the lines.  {\it Both} in the observations and simulations, the majority of the evolution with redshift is in $\mathrm{H}\upalpha$, with a factor 2 and 1.4 larger difference than measured for the stellar continuum in Log($R_\mathrm{eff}$) respectively, and a factor 2 in Log($\Sigma_{1\mathrm{kpc}}$) for the observations at a fixed stellar mass of Log$(M_{*}/\mathrm{M}_{\odot})=9.5$. See Table~\ref{tab:lines_sep} for the measured relations shown as solid lines in the first and third rows. Upper right: The ratio of the $\mathrm{H}\upalpha$ to stellar continuum effective radius in CLEAR, 3D-HST and TNG50. The IllustrisTNG effective radii follow the observational effective radii measurements remarkably well, suggesting the observed redshift evolution towards more extended $\mathrm{H}\upalpha$ profiles of star-forming galaxies at lower redshifts is {\it not} dust-driven (see Section~\ref{dust}).}
	\label{fig:sig1_reff1_sep}
\end{figure*}

\begin{table*}  
  \centering
    \caption{Least-Squares linear fits to the Log$(\mathrm{R}_{\mathrm{eff}})$- and Log$(\Sigma_{1\mathrm{kpc}})$- stellar mass relations for the stellar continuum and \ha~shown in Figure~\ref{fig:sig1_reff1_sep}.
\label{tab:lines_sep}}
  \begin{tabular}{|c|c|c c c c|}
      \cline{3-6}
    \multicolumn{2}{c|}{} & \multicolumn{2}{c|}{CLEAR} & \multicolumn{2}{c|}{3D-HST} \\ 
    \multicolumn{2}{c|}{} & \multicolumn{2}{c|}{$0.22 \lesssim z \lesssim 0.75$} & \multicolumn{2}{c|}{$0.7<z<1.5$} \\ 
    \multicolumn{2}{c|}{} & \multicolumn{2}{c|}{Median Log$(M_{*}/\mathrm{M}_{\odot})=9.47$} & \multicolumn{2}{c|}{Median Log$(M_{*}/\mathrm{M}_{\odot})=9.62$} \\
    \cline{3-6}
    \multicolumn{2}{c|}{} & Gradient & Intercept & Gradient & Intercept \\ \hline
    \multirow{2}{*}{Log($R_{\mathrm{eff}}$)} & Cont & $0.19186\pm0.00005$ & $-1.4272\pm0.0004$   & $0.1936\pm0.0001$ & $-1.538\pm0.001$  \\
    & \ha & $0.32\pm0.02$ & $-2.6\pm0.2$ & $0.233\pm0.006$ & $-1.89\pm0.06$   \\ \hline
    \multirow{2}{*}{Log($\Sigma_{1\mathrm{kpc}}$)} & Cont & $0.7966\pm0.0001$ & $0.726\pm0.009$ & $0.7824\pm0.0002$ & $0.98\pm0.02$  \\
    & \ha & $0.651\pm0.001$ & $2.0\pm0.1$  & $0.655\pm0.002$ & $2.2\pm0.2$ \\ \hline
  \end{tabular}
\end{table*}

The main result of this paper is that star-forming galaxies at lower redshifts have lower \ha-to-stellar continuum surface densities within 1 kpc, but similar \ha-to-stellar continuum effective radii to their high redshift counterparts. This is an intriguing result, for which there are multiple physical explanations. To help rule out some of these physical explanations, we perform a comparison of our observational results to similar, albeit not identical, measurements made to galaxies in the cosmological hydrodynamical simulation TNG50 \citep{Pillepich2019, Nelson2019}, of the IllustrisTNG project\footnote{\url{www.tng-project.org}}. \cite{Pillepich2019} studied the stellar and \ha~disks of thousands of star-forming galaxies in the TNG50 simulation between $0\lesssim z \lesssim 6$. Their measurements provide us with the opportunity to attempt a direct comparison between observational and simulated data.

\subsection{Sizes and comparable galaxy samples from TNG50}
\label{tng_comparable}

To ensure as direct a comparison as possible with simulations, it is imperative that we ensure both observables and sample selection are as unbiased as possible. 

In the case of observables, we use stellar mass and star formation rate (SFR) measurements from the TNG50 results that are comparable to the analagous quantities measured from our observations. However, we do not undertake a true forward modeling of the observational quantities. Instead, following \cite{Pillepich2019}, we use the stellar mass within 30 kiloparsecs from the galaxy centre as our comparable stellar mass. From our observations, we derive SFRs from our grism \ha~emission line measurements, corrected for dust and [\ion{N}{2}] contribution\footnote{Following \cite{Sobral2015}, we scale the \ha~flux down assuming [\ion{N}{2}]/\ha~= 0.25. Our dust correction follows the method outlined in Section 6.3.2 of \cite{Tiley2020} where the \ha~extinction of stars is calculated using the stellar extinctions, $A_{v}$, from the \texttt{FAST} \citep{Kriek2009} catalogs with the \cite{Calzetti1999} dust law.}. \ha~emission traces young stars with lifetimes of approximately 10 Myr \citep{KennicuttJr.1998}. Therefore in TNG50, we choose to compare to the SFR averaged over the last 10 Myr measured within 30 kiloparsecs from the galaxy centre. We compare our stellar continuum effective radii to the half-light radii in the $V$ band measurements made on face-on 2D projections of TNG50 galaxies. Our \ha~effective radii are compared to the half-light radii measurements made on face-on 2D projections of \ha~emission for the same TNG50 galaxies. The size estimates from TNG50 are all circularized; moreover, they represent the intrinsic extents of the simulated galaxies, as the effects of dust are not accounted for.

Next, to ensure our samples of TNG50 galaxies are as directly comparable to our observational samples as possible, we select samples of TNG50 star-forming galaxies at $z=0.5$ and $z=1$ that have the same stellar mass and SFR distributions as the star-forming galaxies in our CLEAR and 3D-HST samples respectively. We do this by performing two-dimensional kernel density estimation, assigning probabilities to TNG50 star-forming galaxies based on how well they follow the observed stellar mass and SFR distributions for their respective redshifts. We draw 1000 samples each containing 100 galaxies that match the stellar mass and SFR distributions of our CLEAR sample from the $z=0.5$ TNG50 snapshot and do the same for the $z=1$ snapshot with respect to the 3D-HST stellar mass and SFR distributions.

\subsection{Results on the comparison to Simulations}

Figure~\ref{fig:sig1_reff1_sep} shows our observational and TNG50 measurements separated by stellar continuum and \ha~in the left panel to help disentangle where most of the redshift evolution occurs. The TNG50 results are binned using the same bins as those used for the observations\footnote{For simplicity, we only use the CLEAR All fields bins for $z=0.5$ in IllustrisTNG.}. Means are calculated for each bin, as we posit that the mean is the statistic most closely related to what the observational stack measurements represent. The shaded regions show these means and $1\sigma$ standard deviation from measurements made for the 1000 samples (see Section~\ref{tng_comparable}). The thick solid lines in the first and third rows of the left panel are least-squares fits to the observational measurements, details for which are given in Table~\ref{tab:lines_sep}. Second and fourth rows show Log($\mathrm{R}_{\mathrm{eff, CLEAR}}$)-Log($\mathrm{R}_{\mathrm{eff, 3D-HST}}$) and Log($\mathrm{\Sigma}_{1\mathrm{kpc, 3D-HST}}$)-Log($\mathrm{\Sigma}_{1\mathrm{kpc, CLEAR}}$) in green over the same stellar mass range using the linear fits respectively. The solid line shows the stellar mass range over which there are measurements from both surveys and the dashed lines extrapolations. The blue line is the same, but for TNG50. The numbers stated in these plots are the least-squares measured slopes of the solid lines.

The most striking result when looking at the observational results by eye is that the majority of the redshift evolution is in H$\upalpha$, not the stellar continuum. This is seen in the larger $y$-separation between the CLEAR and 3D-HST linear fits in the first and third row of the second column compared to the first column. We can see this more quantitatively in the corresponding difference plots in the second and fourth rows. At a fixed stellar mass of Log$(M_{*}/\mathrm{M}_{\odot})=9.5$, the difference between the CLEAR and 3D-HST Log$(R_{\mathrm{eff}})$ measurements is 2 times larger in \ha~than in the stellar continuum. The same level of evolution is seen for Log$(\Sigma_{1\mathrm{kpc}})$. This larger evolution in \ha~is also seen in TNG50 Log$(R_{\mathrm{eff}})$ measurements, with a factor 1.4 times larger difference in \ha~compared to the stellar continuum. It is also evident that this difference rises towards higher stellar masses in the observations, due to the positive slopes in the Log$(R_{\mathrm{eff}})$ and Log$(\Sigma_{1\mathrm{kpc}})$ \ha~differences. This is not true in the simulation, where the stellar continuum and \ha~Log$(R_{\mathrm{eff}})$ differences have the same slope.

Another important takeaway from this comparison is that the TNG50 $R_{\mathrm{eff}}$ measurements in both the stellar continuum and \ha~follow the observational results remarkably well. Even though here the operational definition of sizes is not identical between observed and simulated galaxies (circularized vs. major axis, effective radii measured directly versus GALFIT-based S\'ersic profile fitting). A similarly good level of agreement between TNG50 and observational results at $z\sim1$ has been noticed before in \cite{Nelson2021}, for the locus of the star-forming main sequence and SFR profiles. Here we show that in $R_{\mathrm{eff}}$ for both stellar continuum and H$\upalpha$, this encouraging level of agreement continues down to $z\sim0.5$. 

In the next Section, we discuss how the agreement in stellar continuum and \ha~$R_{\mathrm{eff}}$ measurements between the observations and TNG50 can help us constrain the physical explanation responsible for the more extended \ha~profiles in low redshift star-forming galaxies.

\subsection{Why low redshift Star-forming Galaxies have lower \ha~Central Surface Densities at fixed stellar mass}

\subsubsection{Dust}
\label{dust}
The first possible physical explanation for the lower \ha~central surface densities in star-forming galaxies towards low redshifts is higher dust obscuration in the central regions of these galaxies preferentially attenuating \ha~emission from young stars. A natural consequence of inside-out growth via star formation is that the central regions of galaxies are older than their outer regions. Therefore, the star formation history of the inner region is longer than that of the outer region. Consequently, inner regions of galaxies have a higher degree of stellar mass loss, leading to more efficient enrichment of the interstellar medium (ISM). The central regions of galaxies therefore build up a thicker metal column, leading to increased obscuration of light caused by dust \citep{Wuyts2012}. Higher levels of dust at the centers of galaxies with respect to the outskirts leads to preferential dust attenuation of light in the central regions of galaxies. In the context of a galaxy's light profile (e.g. Figure~\ref{fig:profiles}), this will have the effect of flattening out and extending the light profile in the central regions of galaxies (e.g. \citealt{Nelson2016a}), and in the context of our study, increase $R_{\mathrm{eff}}$ and decrease $\Sigma_{1\mathrm{kpc}}$ measurements. 

However, the stellar continuum and \ha~emission for our galaxies both in CLEAR and 3D-HST are measured within the same wavelength range, meaning the level of dust attenuation for both the stellar continuum and \ha~emission should be the same and should therefore cancel out in any H$\upalpha$-to-stellar continuum ratio measured. The only way the relatively more extended \ha~profiles to the stellar continuum profiles at low redshift can be explained by dust is if there is excess attenuation of \ha~emission relative to the stellar continuum. Many works in the literature have found results for \citep{Calzetti1999,ForsterSchreiber2009,Yoshikawa2010,Mancini2011,Wuyts2011,Wuyts2013, Kashino2013, Kreckel2013,Price2014,Bassett2017, Theios2019,Koyama2019a,Greener2020, Wilman2020,Rodriguez-Munoz2021} and against \citep{Erb2006, Reddy2010} higher dust attenuation of \ha~relative to the stellar continuum. This is thought to be driven by the fact that \ha~traces emission from young stars which can still be enshrouded in stellar birth clouds. The stellar continuum, which is dominated by older stars does not suffer significantly from this additional contribution. Regardless of this excess attenuation of $\mathrm{H}\upalpha$, if the {\it relative} dust attenuation of \ha~versus the stellar continuum is independent of redshift, we should not see a difference in the intercept of the $\Sigma_{1\mathrm{kpc, H}\upalpha}$/$\Sigma_{1\mathrm{kpc, Cont}}$--stellar mass relation with redshift.

A handle on the amount of dust attenuation towards star-forming regions can be directly probed using the Balmer decrement, the ratio of the \ha~to $\mathrm{H}\upbeta$ line flux \citep{Calzetti1997}. At $z\sim1.4$, \cite{Nelson2016a} measured spatially resolved Balmer decrements from HST WFC3 G141 grism spectroscopy as part of the 3D-HST survey (Section~\ref{3dhst}). They found that galaxies with a mean stellar mass of Log$(M_{*}/\mathrm{M}_{\odot})=9.2$ -- which is close to the mean stellar mass of both our CLEAR and 3D-HST samples (Log$(M_{*}/\mathrm{M}_{\odot})\sim9.5$) -- exhibit little dust attenuation at all galactocentric radii. Similarly, at $z\sim2$, \cite{Tacchella2018} found that the highest mass star-forming galaxies (Log$(M_{*}/\mathrm{M}_{\odot})\gtrsim11$) have the most significant dust obscuration, predominantly located at their centers. At low redshift with SDSS-IV MaNGA, \cite{Greener2020} recently showed that for star-forming spiral galaxies with Log$(M_{*}/\mathrm{M}_{\odot})<10.26$, there is $2.3\pm0.2$ times more dust attenuation of the H$\upalpha$-emitting gas than the stellar continuum, but this ratio remains constant with galactocentric radius. These works suggest that at high redshift, there is likely negligible dust attenuation for galaxies with similar stellar masses to our CLEAR and 3D-HST galaxies. As we move towards lower redshifts, there is a factor of $\sim2.3$ excess dust attenuation of the \ha~emission, but it follows a similar spatial distribution to the stellar continuum dust attenuation. The level of light profile extension due to dust should therefore be equivalent for both the stellar continuum and \ha~emission. Hence dust is likely not driving the more extended \ha~profiles relative to stellar continuum profiles at low redshift. This will be verified in our upcoming study on the evolution of spatially resolved Balmer decrements between $0.7\lesssim z\lesssim 1.4$. Future {\it James Webb Space Telescope} (JWST) slitless spectroscopy will also help to firm up the nature of spatially resolved dust as a function of stellar mass and redshift.

The comparison of our observational results to TNG50 in Figure~\ref{fig:sig1_reff1_sep} supports the interpretation that dust is not responsible for the redshift evolution in \ha~profiles. The TNG50 measurements are based on the location and intrinsic properties of the stellar particles and star-forming gas cells, without taking the effects of dust into consideration. In other words, the exact location of all star-forming elements is known: a fraction of them are not hidden by dust like they are when relying on \ha~observations. Therefore, if the intrinsic \ha~and continuum structure of TNG50 star-forming galaxies resemble those of real galaxies, the TNG50 sample act as a no-dust control sample to our observational samples of star-forming galaxies. The TNG50 results therefore inform us what evolution we should expect in the absence of dust or non-evolving dust profile gradients in \ha~versus the stellar continuum. The fact that the TNG50 results follow our effective radii measurements so well further consolidates our conclusion that the observed evolution in \ha~profiles is not due to dust, but something more intrinsic to galaxy evolution that would be captured by a cosmological simulation and {\it not} significantly affect galaxy dust profiles.

\subsubsection{Inside-out Quenching versus Inside-out Growth}
\label{quench_vs_growth}

\begin{figure}
	\centering\includegraphics[width=\columnwidth]{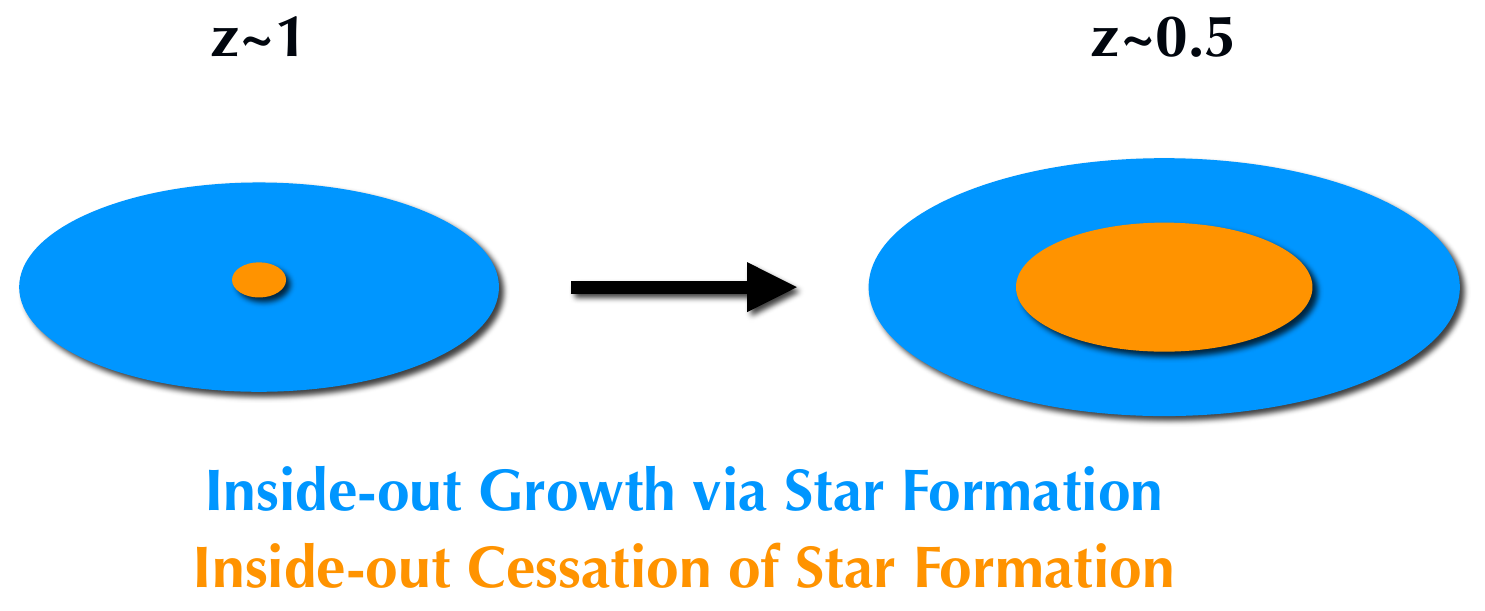}
    \caption{Schematic representation for the natural consequences of inside-out growth via star formation in star-forming galaxies that can explain the redshift evolution seen in our study at $0.5<z<1$ (see Section~\ref{quench_vs_growth} for more details). The inside-out cessation of star formation (orange) will follow behind the inside-out growth (blue), becoming significant at lower redshifts. $\Sigma_{1\mathrm{kpc, H}\upalpha}$ will be more sensitive to the effects of inside-out quenching on the ionized gas (that is traced by \ha).}
    \label{fig:schematic}
\end{figure}

It is a natural consequence of inside-out growth via star formation that the inside-out cessation (or ``quenching") of star formation will follow in its wake. More specifically, if stars form at larger and larger galactocentric radii with time, stars will also age in this direction. That is, the oldest stars will be nearest to the center of the galaxy. This remains true even in the absence of an inside-out quenching mechanism being present. It is worth noting however, that in TNG50, there is no inside-out quenching without AGN feedback \citep{Nelson2021}. The effects of these two physical processes with their varying degrees between $0.5 \lesssim z \lesssim 1$ can successfully explain the results presented in this paper. We show a schematic representation in Figure~\ref{fig:schematic} to help visualize our explanation. This picture demands that the radius within which inside-out quenching occurs is always smaller than the radius within which inside-out growth via star formation occurs. Because the stellar continuum includes light from all stars and \ha~includes emission predominantly from young stars, this dictates that $R_{\mathrm{eff, H}\upalpha}$/$R_{\mathrm{eff, Cont}}>1$ and $\Sigma_{r\mathrm{, H}\upalpha}$/$\Sigma_{r\mathrm{, Cont}}<1$ always. Within our errors, we see this is true for CLEAR, 3D-HST and KMOS$^{3\mathrm{D}}$ (Figures~\ref{fig:comp} \& \ref{fig:sigma_all}). Because the stellar continuum encodes the integrated star formation history of a galaxy, the stellar continuum of an average $z\sim1$ star-forming galaxy will have a higher contribution of younger stellar populations compared to the stellar continuum of an average $z\sim0.5$ star-forming galaxy. This ensures $R_{\mathrm{eff, H}\upalpha}$/$R_{\mathrm{eff, Cont}}\rightarrow1$ and $\Sigma_{r\mathrm{, H}\upalpha}$/$\Sigma_{r\mathrm{, Cont}}\rightarrow1$ towards higher redshifts. This is most apparent when comparing CLEAR and 3D-HST in Figure~\ref{fig:comp}. 

Because star formation is more prevalent at higher redshifts and declines towards lower redshifts \citep{Madau&Dickinson2014,Whitaker2014a}, the effects of inside-out quenching on galaxy light profiles will be more apparent on spatially resolved population studies of star-forming galaxies at lower redshifts. Between $z\sim1$ and $z\sim0.5$, inside-out quenching becomes increasingly significant. Consequently, its effects on the light profiles of star-forming galaxies becomes more apparent in spatially resolved measurements. \ha~profiles are more sensitive to the significant depletion of gas within the inside-out quenching radius, because they trace ionized gas emission. Furthermore, because \ha~emission has a short lifetime compared to the stellar continuum, changes to \ha~profiles will be more easily measurable between $0.5 \lesssim z \lesssim 1$. It is therefore perhaps unsurprising that we measure a significantly larger extension in $R
_{\mathrm{eff}}$ likely due to inside-out growth and drop in surface density within 1 kpc likely due to inside-out quenching in \ha~profiles than stellar continuum profiles between $0.5 \lesssim z \lesssim 1$ (see Figure~\ref{fig:sig1_reff1_sep}). 

From $z\sim1$ to $z\sim0.5$, the average stellar continuum of a star-forming galaxy gains an increasing contribution from older stars that will reside predominantly near the center of star-forming galaxies. These older stars will be less obscured by gas due to the gas depletion that has occurred in the region during the same period. In such a scenario, one would expect that the stellar continuum becomes more compact towards lower redshifts for star-forming galaxies, since the contrast between the bulge and disk increase in favour of the bulge. Since this is not what we measure (see Figure~\ref{fig:sig1_reff1_sep}), it tells us that the net effect on the stellar continuum profile of typical star-forming galaxies from $z\sim1$ to $z\sim0.5$ is that of inside-out growth via star formation. In other words, there is competition between the increasing brightness of the bulge and more luminous, young stars at large galactocentric radii for dominance of the stellar continuum profile at low redshift. Ultimately, the more luminous, younger stars at large galactocentric radii ensure the overall effect on the stellar continuum profile is one of extension and not contraction, albeit less significant than the extension measured in \ha~(first column versus second column in the largest panel of Figure~\ref{fig:sig1_reff1_sep}).

A similar picture to the one presented above was also proposed as an explanation for the results presented in \cite{Azzollini2009} which are remarkably similar to ours. They conducted an analagous study to ours using 270 Log$(M_{*}/\mathrm{M}_{\odot})\sim10$ galaxies at $ 0\lesssim z \lesssim 1$, but used the near ultraviolet ($NUV$) as their tracer for ongoing star formation. The $NUV$ is dominated by flux from O, B, A and F stars that have lifetimes of $\lesssim 100$~Myr \citep{Kennicutt1998,Kennicutt2012} and so traces star formation occurring on longer timescales than the star formation traced by \ha~(10 Myr). It is also well known that the ultraviolet is very sensitive to dust attenuation \citep{Kennicutt2012} and so can be more easily obscured than \ha~emission. 

Neverthless, \cite{Azzollini2009} found that after dust correction, SFR surface density (traced by the $NUV$) decreases from $z\sim1$ to $z\sim0$, with the greatest decrease occurring at galactocentric radii $\lesssim 2.5$~kpc. The largest difference in stellar mass build up at $0 \lesssim z \lesssim 1$ is measured at $1.5-2$~kpc from the centre, suggestive of a decline in star formation activity within $\leqslant1.5-2$~kpc for star-forming galaxies at $z\sim0$. The SFR surface density is positively correlated with the gas surface density as per the Kennicutt-Schmidt law \citep{Schmidt1959,Kennicutt1998,Kennicutt2012}. \ha~emission traces the gas ionized by the UV radiation from young stars. Therefore a fall in the \ha~stellar mass surface density within 1 kpc at fixed stellar mass from $z\sim1$ to $z\sim0.5$ in our work can be interpreted as a fall in star formation activity due to the exhaustion of gas within a a galactocentric radius of 1 kpc.

Furthermore, our results are consistent with those of \cite{Tacchella2015} who showed that star-forming galaxies with similar stellar masses to our CLEAR and 3D-HST galaxies at $z\sim2.2$ also exhibit centrally suppressed SFR surface densities when compared to their stellar mass surface densities. These authors put forward a picture in which the bulges and outer regions of galaxies are built concurrently, with a `compaction' scenario responsible for the high central surface mass densities (see also \citealt{Tacchella2018}). They also noted the often irregular morphologies of the SFR spatial distributions, with bright clumps at large galactocentric radii. Such morphologies are also prevalent at $z\sim0.5$ in our work (e.g. GS4 29845, GS 49441 and GN 22285 in Figures \ref{fig:eyecandy1} \& \ref{fig:eyecandy2}).

The results presented in \cite{Azzollini2009} and \cite{Tacchella2015} provide us with independent verifications that irrespective of the star formation tracers used and when dust obscuration is insignificant (see Section~\ref{dust}) or corrected for, spatially resolved \ha/{\it NUV} and stellar continuum morphological measurements are measuring the same physical processes acting in typical star-forming galaxies as a function of redshift. Because these physical processes are a product of how inside-out growth operates and inside-out growth is expected in our canonical model of galaxy formation (see Section~\ref{sec:intro}), it is perhaps unsuprising that the TNG50 simulation captured its effect on the \ha~and stellar continuum $R_{\mathrm{eff}}$ so well.

\section{Summary}
\label{summary}

Using deep HST WFC3 G102 grism spectroscopy of star-forming galaxies from the CLEAR survey (Section~\ref{CLEAR}), we have studied spatially resolved star formation using \ha~emission line maps at $z\sim0.5$ in this paper. The CLEAR data set made it possible to study spatially resolved star formation over a wide range in redshift for the first time, by comparing our results to analogous studies conducted at $z\sim1$ and $z\sim1.7$ with the 3D-HST (Section~\ref{3dhst}) and KMOS$^{\mathrm{{3D}}}$ (Section~\ref{kmos3d}) surveys.

We processed, stacked, and analysed our data using the same methodology as 3D-HST, but deferred to improved methods where our more sophisticated software allowed for them.

Our main conclusions are as follows:

\begin{enumerate}
    \item The \ha~effective radius for star-forming galaxies with  Log$(M_{*}/\mathrm{M}_{\odot})\geqslant8.96$ at $z\sim0.5$ is $1.2\pm0.1$ times larger than the effective radius of their stellar continuum (Figure~\ref{fig:ms}, second panel of Figure~\ref{fig:comp} and Table~\ref{tab:means}). Interpreting \ha~emission as star formation, this implies star-forming galaxies at $z\sim0.5$ are growing inside-out via star formation.
    
    \item The \ha~to stellar continuum effective radius ratio measured at $z\sim0.5$ from CLEAR agrees within $1\upsigma$ with the values measured at $z\sim1$ and $z\sim1.7$ from the 3D-HST and KMOS$^{\mathrm{{3D}}}$ surveys (second panel of Figure~\ref{fig:comp} and Table~\ref{tab:means}). This implies there is no change in the pace of inside-out growth via star formation with redshift.
    
    \item By removing the S\'ersic index -- effective radius degeneracy with the stellar mass surface density within one kiloparsec, $\Sigma_{1\mathrm{kpc}}$, we unveil a redshift evolution in the \ha~to continuum $\Sigma_{1\mathrm{kpc}}$ ratio (bottom panel of Figure~\ref{fig:comp},  Tables~\ref{tab:means} and \ref{tab:lines}). Star-forming galaxies at $z\sim0.5$ have a $(19\pm2)\%$ lower surface density in \ha~relative to their stellar continuum within 1 kpc at a fixed stellar mass of Log$(M_{*}/\mathrm{M}_{\odot})=9.5$.
    
    \item The $\Sigma_{1\mathrm{kpc, H}\upalpha}$/$\Sigma_{1\mathrm{kpc, Cont}}$--stellar mass relation has a linearly declining relation with stellar mass at all redshifts, with a non-evolving slope between $0.5<z<1$ (bottom panel of Figure~\ref{fig:comp} and Table~\ref{tab:lines}).

    \item We advocate for the use of $\Sigma_{1\mathrm{kpc}}$ over other morphology parameters for similar studies to ours in the future (Section~\ref{sig1_advantages}). The reduced scatter due to the breaking of the S\'ersic index -- effective radius degeneracy increases precision in $\Sigma_{1\mathrm{kpc}}$ measurements compared to morphology parameters that rely on S\'ersic index or effective radius alone. Furthermore, its reliance on the inner, most highest signal-to-noise ratio region -- and therefore most well-determined part of the stellar continuum and \ha~light profiles (Figure~\ref{fig:profiles}) -- increases its sensitivity towards radial trends that are otherwise washed out at large radii (Figure~\ref{fig:sigma_all}).
    
    \item We find that most of the redshift evolution seen in the $\Sigma_{1\mathrm{kpc, H}\upalpha}$/$\Sigma_{1\mathrm{kpc, Cont}}$--stellar mass relation is driven by changes in \ha~profiles, with a factor 2 larger drop in Log($\Sigma_{1\mathrm{kpc, H}\upalpha}$) than Log($\Sigma_{1\mathrm{kpc, Cont}}$) from $z\sim1$ to $z\sim0.5$ (Figure~\ref{fig:sig1_reff1_sep}).

    \item By comparing our observational results to analogous measurements of galaxies from the TNG50 hydrodynamical simulation in \cite{Pillepich2019}, we find good agreement in $R_{\mathrm{eff}}$ measurements. Using this comparison, we rule out dust or any physical processes that would alter the shape of galaxy dust profiles as a likely driver for the observed redshift evolution (Section~\ref{dust}). 
    
    \item Our results and similar observational results in the literature are consistent with the picture in which the inside-out cessation (or ``quenching") of star formation that naturally follows in the wake of inside-out growth via star formation significantly reduces $\Sigma_{1\mathrm{kpc, H}\upalpha}$ from $z\sim1$ to $z\sim0.5$ (Section~\ref{quench_vs_growth}).

\end{enumerate}

In our follow up paper, we will verify whether there is any redshift evolution in the dust profiles of star-forming galaxies by measuring spatially resolved Balmer decrements using the CLEAR dataset at $z\sim0.7$ and comparing it to the analagous study done with the 3D-HST survey at  $z\sim1.4$ \citep{Nelson2016a}. This will provide us with a quantitative dust correction and unveil whether or not dust attenuation towards \ion{H}{2} regions evolves with redshift.

\vspace{1pt}

\acknowledgments

\noindent JM thanks Matteo Fossati, Erica J. Nelson and Annalisa Pillepich for providing the KMOS$^{\mathrm{3D}}$, 3D-HST and TNG50 measurements respectively, which made the comparisons to \cite{Wilman2020}, \cite{Nelson2015} and \cite{Pillepich2019} possible. JM also thanks Robert Kennicutt for helpful discussions regarding the results in this paper. This work is based on data
obtained from the Hubble Space Telescope through program number
GO-14227.  Support for Program number GO-14227 was provided by NASA
through a grant from the Space Telescope Science Institute, which is
operated by the Association of Universities for Research in Astronomy,
Incorporated, under NASA contract NAS5-26555.  This work is supported
in part by the National Science Foundation through grant AST 1614668.
VEC acknowledges support from the NASA Headquarters under the Future
Investigators in NASA Earth and Space Science and Technology (FINESST)
award 19-ASTRO19-0122. This work was supported
in part by NASA contract  NNG16PJ33C, the Studying Cosmic Dawn with
WFIRST Science Investigation Team. JM, CP and VEC are also grateful for the support from the George P. and Cynthia Woods Mitchell Institute for Fundamental Physics and Astronomy at Texas A\&M University.

\software{This research made use of \textsc{Astropy}, a community-developed core Python package for Astronomy \citep{TheAstropyCollaboration2018}. The python packages \textsc{Matplotlib} \citep{Hunter2007}, \textsc{Numpy} \citep{VanDerWalt2011}, and \textsc{Scipy} \citep{scipy} were also extensively used. Parts of the results in this work make use of the colormaps in the \textsc{CMasher} \citep{VanderVelden2020} package.}

\facilities{ {\it HST}\ (NASA/ESA)}

\appendix
\label{appendix}

\begin{figure*}

	\centering\includegraphics[width=0.5\textwidth]{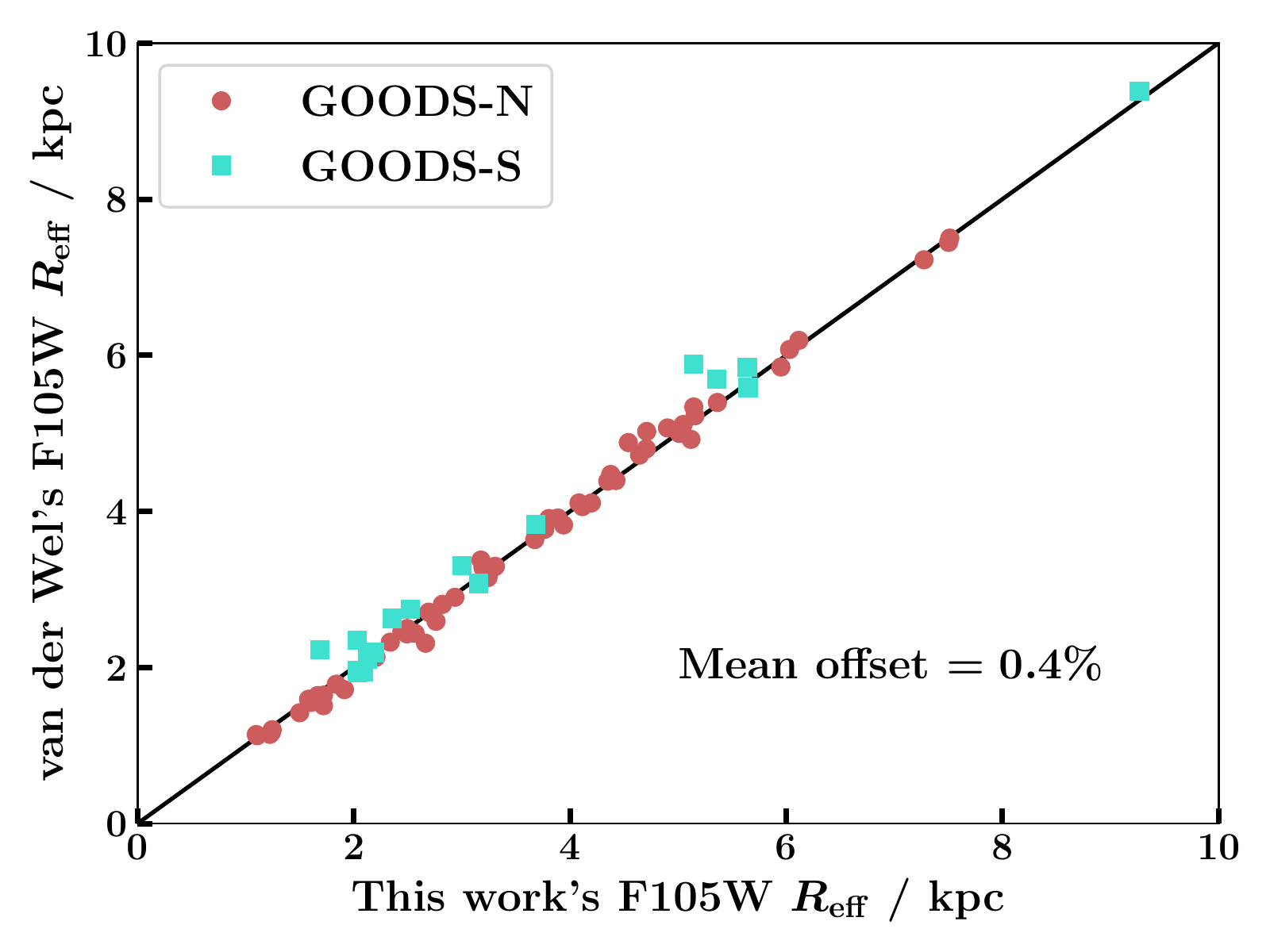}
    \caption{Level of agreement with results from \cite{VanderWel2012} for the F105W effective radii of the same galaxies from individual fits shown in Figure~\ref{fig:ms}. Only measurements that had good quality fits (flag value = 0) in the \cite{VanderWel2012} catalogs are shown. Solid black line indicates the position of one-to-one agreement between the two size determination methods. Our sizes are on average $0.4\%$ smaller than those of \cite{VanderWel2012}.}
    \label{fig:size_size}
\end{figure*}


\bibliography{library_grizli}{}
\bibliographystyle{aasjournal}



\end{document}